\documentclass[submission, Phys]{SciPost}
\usepackage{amsmath,braket,mathdots,mathtools,amssymb,xcolor,cancel}
\usepackage{graphicx}
\usepackage{hyperref}
\usepackage{lipsum}
\usepackage{bbm}

\newcommand{\phdag}{\vphantom{\dagger}}

\newcommand{\sech}{\mathrm{sech \,}}

\def\be{\begin{equation}}
\def\ee{\end{equation}}
\def\bea{\begin{eqnarray}}
\def\eea{\end{eqnarray}}
\def\nn{\nonumber\\}
\def\fr#1{(\ref{#1})}
\def\secfix#1{\texorpdfstring{#1}{Lg} }
\begin{document}

\begin{center}
{\Large\bf 
On the low-energy description for tunnel-coupled one-dimensional Bose
gases} 
\end{center}
\begin{center}
Yuri D. van Nieuwkerk${}^\star$ and Fabian H. L. Essler,
\end{center}
\begin{center}
The Rudolf Peierls Centre for Theoretical Physics,\\
Oxford University, Oxford OX1 3PU, UK\\
${}^\star${\small \sf yuri.vannieuwkerk@physics.ox.ac.uk}
\end{center}
\date{\today}

\section*{Abstract}
We consider a model of two tunnel-coupled one-dimensional Bose gases
with hard-wall boundary conditions. Bosonizing the model and retaining
only the most relevant interactions leads to a decoupled theory
consisting of a quantum sine-Gordon model and a free boson, describing
respectively the antisymmetric and symmetric combinations of the phase
fields. We go beyond this description by retaining the perturbation
with the next smallest scaling dimension. This perturbation carries
conformal spin and couples the two sectors. We carry out a detailed
investigation of the effects of this coupling on the non-equilibrium
dynamics of the model. We focus in particular on the role played by
spatial inhomogeneities in the initial state in a quantum quench setup. 

\vspace{10pt}
\noindent\rule{\textwidth}{1pt}
\tableofcontents\thispagestyle{fancy}
\noindent\rule{\textwidth}{1pt}
\vspace{10pt}

\section{Introduction}
\label{sec:introduction}

The study of one-dimensional quantum many-body systems out of equilibrium has seen great progress in the past decades. Long-standing questions concerning the equilibration of observables, spreading of correlations and entanglement, and the emergence of statistical mechanics from microscopics have been successfully tackled using a range of innovative theoretical ideas \cite{Rigol2007,CalabreseCardy2007,Polkovnikov2011,EFreview,Vidmar2016,DAlessio2016,Gogolin2016,CalabreseCardy2016,CalabreseRev2018}, whilst spectacular advances in the ability to realize archetypical one-dimensional quantum many-body sytems using cold atoms \cite{Ketterle2001,Greiner2001,Kinoshita2004,Schumm2005,Hofferberth2007} have made it possible to test many of these theoretical developments using tabletop experiments \cite{Trotzky2012,Cheneau2012,Gring2012,Langen2013,Langen2015,Kaufman2016}. However, such experimental engineering of quantum many-body Hamiltonians relies on certain assumptions to make the experiments map onto a model of physical interest. These assumptions often include having a low energy density, at which an effective low-energy theory holds, and translational invariance, which can generally simplify the problem and specifically play an important role in the integrability of the low-energy theory. When studying non-equilibrium problems in finite quantum many-body systems, these two assumptions are sometimes brought into question.

We here study a situation where both the successes and challenges
described above are clearly present: we consider pairs of
tunnel-coupled, elongated Bose gases, as realized in the Vienna
experiments
\cite{Schumm2005,Hofferberth2007,Gring2012,Kuhnert2013,Smith2013,Langen2013,Langen2015,Schweigler2017,Pigneur2017}. An
interesting feature of these experiments is that in certain limits,
density measurements after matter-wave interference
\cite{Ketterle1997,Schumm2005} correspond to projective von Neumann
measurements of the relative phase field \cite{Nieuwkerk2018}. This
allows for the reconstruction of full distribution functions of
quantum mechanical observables
\cite{Kuhnert2013,Smith2013,Schweigler2017}, which is of considerable
theoretical interest
\cite{cd-07,Imambekov2007,lp-08,ia-13,sk-13,e-13,k-14,sp-17,CoEG17,nr-17,hb-17,lddz-15,bpc-18,Groha18,Collura20}
in general. In the case at hand, situations without tunnel-coupling
can be modelled by a two-component Luttinger liquid
\cite{Haldane1981,Bistritzer2007}. This description in terms of a
quadratic quantum critical model has yielded theoretical results for 
the full fluctuation statistics of the relative phase field
\cite{Imambekov2007,Kitagawa2010,Kitagawa2011} which show a satisfying
match with experimental results \cite{Gring2012,Langen2015}.

Our interest lies in the effect of a finite tunnel barrier between the
gases \cite{Albiez2005,Gati2006,Levy2007,Hofferberth2007}. This
introduces a relevant perturbation and at sufficiently low energies
leads to a decoupled theory of a Luttinger liquid describing the
symmetric combination of Bose gas phases (``symmetric sector'') and a sine-Gordon
model \cite{Gritsev2007} describing the relative phase (``antisymmetric sector'').
The sine-Gordon model is of great theoretical importance as it is
an exactly solvable, Lorentz invariant quantum field theory that
exhibits a rich range of physical phenomena like dynamical mass
generation and topological excitations and moreover has important
applications to electronic degrees of freedom in solids
\cite{EsslerKonik2005}. Its behaviour out of equilibrium has received
a lot of attention in the past decade. To be able to study dynamics,
the very weakly interacting limit is amenable to a simple harmonic
approximation \cite{Iucci2010,Foini2015,Foini2017}, while the free
fermion point can also be used to obtain exact results
\cite{Iucci2010}. Integrability-based methods were used in 
Refs.~\cite{Bertini2014,Schuricht2017,Horvath2017,Horvath2018a} to
study quenches from ``integrable'' initial states, whereas
semiclassical methods \cite{Kormos2016,Pascu2017} were applied to the
study of the time-dependence of one and two-point functions as well as
the probability distribution of the phase. The truncated conformal space
approach\cite{James2018} was employed in Ref.~\cite{Kukuljan2018} to
analyse the time evolution of two and four-point functions after a
quantum quench. A first litmus test for the experimental realization
of the sine-Gordon model using split Bose gas experiments was
performed in an equilibrium situation: high order equilibirum
correlation functions extracted from projective phase measurements
in the classical limit have been found to agree well with classical
field simulations \cite{Schweigler2017}. A quantum many-body
treatment of some such correlation functions is available as well,
with possible generalizations to non-equilibrium situations
\cite{Sotiriadis2018}. For non-equilibrium initial conditions,
however, experimental studies
\cite{Pigneur2017,Pigneur2019thesis,schweigler2019correlations} have
shown puzzling behaviour: when preparing two elongated Bose gases with
an initial phase difference, applying a tunnel-coupling between them
sets Josephson oscillations of density and phase in motion. These
oscillations show a rapid damping, accompanied by a narrowing of the
distribution function of the phase. To date, no satisfying theoretical
explanation of this damping is known \cite{Pigneur2018}. The damping
seems incompatable with a description in terms of a translationally
invariant sine-Gordon model, which fails to provide a mechanism for
the observed strong and rapid damping in both a self-consistent
harmonic treatment \cite{Nieuwkerk2018b} and in a combination of
truncated Wigner and truncated conformal space approaches
\cite{Horvath2018}. 

In this work, we go beyond previous studies in two important ways:
\begin{enumerate}
\item{}
We take into account the next most relevant perturbation at low
energies. This perturbation induces an interaction between the
symmetric and antisymmetric sectors.
\item{}
We drop the assumption of translational invariance. To this end
we place the model in a hard-wall box geometry and consider
inhomogeneous initial conditions.
\end{enumerate}
Our strategy is to treat the resulting two-component model in the
self-consistent time-dependent approximation (SCTDHA) as described in
\cite{Nieuwkerk2018b}. We consider the dynamics after initializing the
system in a state in which the sectors are uncorrelated and observe how
the new coupling term causes correlations between the two sectors to
develop over time. In addition to this, energy starts to oscillate
between the sectors. Depending on the initial density profile
imprinted on the gas, Josephson oscillations of density and phase are
affected by the presence of the additional term, showing modulations
of the amplitude that differ from the ones observed in the SCTDHA
treatment of isolated sine-Gordon dynamics \cite{Nieuwkerk2018b}.

This paper is organized as follows. In
Sec. \ref{sec:tunnel_coupled_bose_gases_in_a_hard_wall_box}, we
introduce the low-energy effective theory in a box geometry, the
additional interaction term and the observable relevant for
experiment. We also establish some notational conventions. In
Sec. \ref{sec:self_consistent_time_dependent_harmonia_approxiation},
we recapitulate the self-constistent time-dependent harmonic
approximation as well as the framework to compute observables and some
important distribution functions. In
Sec. \ref{sec:results_for_experimentally_relevant_initial_states}, we
apply our formalism to an initial state which is commonly used in the
literature, and present results on energy flow and growth of
correlations between the sectors, along with the effect on Josephson
oscillations, due to the additional interaction
term. Sec. \ref{sec:conclusion} summarizes our conclusions and
discusses questions for further study.

\section{Tunnel-coupled bose gases in a hard-wall box} 
\label{sec:tunnel_coupled_bose_gases_in_a_hard_wall_box}
An appropriate model for the experiments carried out by the Vienna group
is an interacting Bose gas confined in three-dimensional space by a
tight harmonic potential in the $z$-direction, a double-well potential
$V_\perp(y)$ in the $y$-direction and a shallow harmonic potential in the
$x$-direction. We will refer to the $x$-direction as
\textit{longitudinal}, and to the remaining directions as
\textit{transverse}. To simplify the problem, we take the 
longitudinal potential to be an infinite square well
\begin{align}
V_{||}(x) = \begin{cases}
	0 &\text{ if } 0 <x <L\ ,\\
	\infty  &\text{ otherwise.}
\end{cases} 
\end{align} 
Just like a shallow harmonic potential this breaks translational
invariance in the longitudinal direction, but has the advantage to be
considerably simpler to analyze. Our starting point is thus the
following Hamiltonian 
\begin{align}
H_{\mathrm{3d}}=\int dx\ dy\ dz\ \bigg\{\Psi^\dagger(x,y,z)\left[-\frac{\nabla^2}{2m}+ V_{||}(x) + V_\perp(y)+\frac{m\omega_{z}^2}{2}z^2 \right]\Psi(x,y,z)\notag \\
+c\big(\Psi^\dagger(x,y,z)\big)^2\big(\Psi(x,y,z)\big)^2\bigg\}\ ,
\end{align}
where $\Psi(x,y,z)$ are complex Bose fields obeying the usual bosonic
commutation relations. 

\subsection{Low-energy effective theory} 
\label{sub:low_energy_effective_theory}
In situations where the transverse potentials are sufficiently tight,
the dynamics in the $y$- and $z$-directions can be integrated out, in
a way analogous to Ref. \cite{Olshanii1998}. Details of this procedure
will be reported elsewhere \cite{TBP}.
Projecting to the lowest
two states of the transverse potential, and taking appropriate linear
combinations of these, we obtain a Hamiltonian for two species of
bosons, $\Psi_{1,2}$, which are approximately localized in wells $1$
and $2$: 
\begin{align}
H_{\mathrm{1d}} = \int_{0}^{L} dx\,\Bigg[ \sum_{j=1,2} \frac{1}{2m} \partial_{x} \Psi^{\dagger}_{j}(x) \partial_{x}
  \Psi_{j}(x) &+ \sum_{j,k,l,m=1,2} \Gamma_{jklm}\,\Psi^{\dagger}_{j}(x)\Psi^{\dagger}_{k}(x)\Psi^{\phdag}_{l}(x)\Psi^{\phdag}_{m}(x) \notag\\
  &- \left( T_{\perp} \Psi^{\dagger}_{1}(x)\Psi_{2}(x) + \mathrm{h.c.} \right)  \Bigg] .
\label{eq:micr_1d_ham}
\end{align}
Here the Bose fields $\Psi(x)$ have commutation relations
$\left[ \Psi_{i}^{\phdag}(x),  \Psi_{j}^{\dagger}(x^{\prime})   \right] = \delta_{i,j}
\delta(x-x^{\prime})$. The two Bose gases are coupled by a tunnelling
term as well as contact interactions. The corresponding coupling
constants $\Gamma_{jklm}$ follow from the details of the low-energy
projection \cite{TBP}. For our purposes, we will assume the diagonal
elements to be equal to the usual Lieb-Liniger interaction constant,
$\Gamma_{jjjj} = g \, \forall \, j$. Hard-wall boundary conditions are
imposed by restricting our problem to states $\ket{\Phi}$ where the density at
the boundary has a vanishing eigenvalue: 
\begin{align}
\Psi^{\dagger}_{j}(L)\Psi^{\phdag}_{j}(L)\ket{\Phi} = \Psi^{\dagger}_{j}(0)\Psi^{\phdag}_{j}(0)\ket{\Phi} = 0. \label{eq:hard_wall_cond}
\end{align}
The one-dimensional model \fr{eq:micr_1d_ham} gives an accurate
description of the full theory $H_{\rm 3d}$ at energies that are small
compared to the energy $E_{\perp,2}$ of the second excited state of
the transverse confining potential. In the actual experiments this is
a large energy scale. The physics of interest occurs at energies that
are small compared to $v/\xi\ll E_{\perp,2}$, where $\xi$ is the
coherence length and $v$ the speed of sound. This enables us to make a
second low-energy projection by employing bosonization \cite{Haldane1981}
\begin{align}
\Psi^{\dagger}_{j}(x) \sim  \sqrt{\rho_{0} + \partial_{x}
  \theta_{j}/\pi} \, e^{-i \phi_{j}(x)} \sum_{m=-\infty}^{\infty}
B_{m} e^{2 m i (x \pi \rho_{0}+ \theta_{j})}
\ . \label{eq:bosonization_identity} 
\end{align}
This provides a low-energy description of (\ref{eq:micr_1d_ham}) in
terms of phase fields $\phi_{j}$ and $\theta_{j}$ with a cutoff length
scale set by the coherence length of the gases, which for weak
interactions is given by $\xi = \pi/m v$ (the sound velocity $v$ is
defined below). The hard-wall condition is encoded in the boundary
conditions of the $\theta$-fields in a way that is described in
Sec. \ref{sub:mode_expansions_for_H0}. Let us first consider the case
where interactions and tunnelling between the two gases are absent,
meaning that both $T_{\perp}$ and the non-diagonal elements of
$\Gamma$ are zero. This leaves us with two Lieb-Liniger models in a
hard-wall box, with interaction strength $g$. Under the mapping
(\ref{eq:bosonization_identity}), the low-energy physics of this model
maps to a pair of Luttinger liquids 
\begin{align}
H_{j} &= \frac{v}{2 \pi} \int_{0}^{L} dx \left[ \frac{1}{K} \left(
  \partial_{x} \theta_{j}(x) \right)^2 + K \left( \partial_{x}
  \phi_{j}(x) \right)^2 \right] , \;\;\;\;
j=s,a. \label{eq:Luttinger_Liquids} 
\end{align}
Here we have defined (anti)symmetric combinations of the phase fields
by
\be
\phi_{s/a} = \phi_{1} \pm \phi_{2}\ ,\quad
\partial_{x}\theta_{s/a} = \frac{\partial_{x}\theta_{1} \pm
\partial_{x}\theta_{2}}{2}\ .
\ee
These fields are compact $\phi = \phi + 2 \pi$, $\theta = \theta +
\pi$ and fulfil commutation relations
\begin{align}
\left[ \partial_{x} \theta_{j}(x), \phi_{l}(y) \right] = i \pi
\delta_{j,l} \delta(x - y).
\end{align}
This implies that the canonically conjugate fields to $\phi_{s/a}$ are given by
\begin{align}
\Pi_{j}(x) \equiv \frac{\partial_{x} \theta_{j}(x)}{\pi}\ .
\end{align}
For weak interactions, the sound velocity $v$ and Luttinger parameter
$K$ are related to the parameters in the Lieb-Liniger model in a
simple way \cite{Cazalilla2004} 
\begin{align}
v &= \frac{\rho_{0}}{m} \sqrt{\gamma} \left( 1 -
\frac{\sqrt{\gamma}}{2 \pi}\right)^{1/2}, \quad K =
\frac{\pi}{2\sqrt{\gamma}} \left( 1 - \frac{\sqrt{\gamma}}{2
  \pi}\right)^{-1/2}. \label{eq:micr_pars}
\end{align}
Here $\gamma = m g/\rho_{0}$ is the dimensionless interaction strength
and $\rho_{0}$ the average density of each of the two Bose gases.

In the next step we take into account the tunnelling term in
\fr{eq:micr_1d_ham} as well as ``off-diagonal'' interaction terms
proportional to $\Gamma_{ijkl}$ with not all indices being
equal. These introduce relevant perturbations (in the renormalization
group sense) with respect to the critical Hamiltonian
(\ref{eq:Luttinger_Liquids}). Inserting the bosonization identity 
(\ref{eq:bosonization_identity}) and assuming $\Gamma$ to be real,
permutation symmetric and symmetric under $ 1 \leftrightarrow 2$, we
find that the perturbations with the lowest scaling dimensions can be
written in the form
\begin{align}
H_{\perp} = - 2t_{\perp} \int_{0}^{L} dx \, \left[\rho_{0} + \sigma \Pi_{s}(x) \right] \cos \phi_{a}(x) \ , \label{eq:mixing_hamiltonian}
\end{align}
where $t_\perp$ and $\sigma$ depend on the microscopic parameters in
\fr{eq:micr_1d_ham}. Importantly, the two terms in
(\ref{eq:mixing_hamiltonian}) get generated independently and we
therefore will treat $t_\perp$ and $\sigma$ as independent
phenomenological parameters in the following.
The Hamiltonian $H_s+H_a+H_\perp$ should be viewed as the
result of integrating out high energy degrees of freedom in a
renormalization group sense. As $t_\perp$ grows much faster than
$t_\perp\sigma$ under the renormalization group it would be unphysical
to consider very large values of $\sigma$. We have therefore
restricted the numerical analyses reported below to the range
$0\leq\sigma\leq 2$. In addition to
\fr{eq:mixing_hamiltonian} there are other perturbations with higher
scaling dimensions. Their systematic derivation as well as an analysis
of their effects will be presented elsewhere \cite{TBP}. In the
case $\sigma=0$ the full low-energy theory decouples into symmetric
and antisymmetric sectors $H=H_s+H'_a$, where $H'_a$ is the
Hamiltonian of a quantum sine-Gordon model \cite{Gritsev2007} 
\be
H'_a=\frac{v}{2 \pi} \int_{0}^{L} dx \left[ \frac{1}{K} \left(
  \partial_{x} \theta_{a}(x) \right)^2 + K \left( \partial_{x}
  \phi_{a}(x) \right)^2 \right]
- 2t_{\perp}\rho_{0}\int_{0}^{L} dx \, \cos \phi_{a}(x).
\ee
The non-equilibrium dynamics of this model was analyzed for the
translationally invariant case in the framework of a SCTDHA in our
recent work \cite{Nieuwkerk2018b}. The additional $\sigma$-term in
(\ref{eq:mixing_hamiltonian}) couples the sine-Gordon model to the
Luttinger liquid Hamiltonian $H_{s}$. In the following we extend the
analysis \cite{Nieuwkerk2018b} to
\begin{align}
  H = H_{a} + H_{s}+H_{\perp}.
\label{eq:full_hamiltonian}
\end{align}

\subsection{Time-of-flight measurements} 
\label{sub:time_of_flight_measurements}
In the Vienna experiments
\cite{Schumm2005,Hofferberth2007,Hofferberth2008,Gring2012,Langen2013,Langen2015,Schaff2015,Pigneur2017,Schweigler2017} measurements are performed by turning off the trapping
potential at some time $t_{0}$, letting the gas expand freely and
imaging the three-dimensional boson density after a time-of-flight
$t_{1}$. The outcome of each such ``single-shot'' measurement is
determined by the eigenvalues $e^{\frac{i}{2}  \varphi_{a,s}(x,t)}$ of
the bosonic vertex operators $e^{\frac{i}{2}\phi_{a,s}(x,t_{0})}$ 
\cite{Imambekov2007,Nieuwkerk2018}. As shown in \cite{Nieuwkerk2018}, 
the result of a single measurement of the boson density after a
time-of-flight $t_{1}$ in the regime relevant for the Vienna
experiments can be well approximated by  
\begin{align}
&\varrho_{\mathrm{tof}}(x,\vec{r},t_{1},t_0)
\simeq \rho_{0} \Big| f(\vec{r},t_{1})\Big|^{2} \times \notag \\
& \Big|\int dx^{\prime}\,
G(x-x^{\prime},t_{1})\Big[ e^{i \frac{m}{2t_{1}}\vec{r} \cdot \vec{d}}
  e^{\frac{i}{2}\left( \varphi_{s}(x^{\prime},t_0)+ \varphi_{a}(x^{\prime},t_0) \right)} + e^{-i    \frac{m}{2t_{1}}\vec{r} \cdot \vec{d}} e^{\frac{i}{2}\left(
    \varphi_{s}(x^{\prime},t_0) - \varphi_{a}(x^{\prime},t_0) \right) }  \Big] \Big|^2 .
\label{eq:density_tof_bos_longitudinal_approx}
\end{align}
Here $\vec{d}$ is the distance between the minima of the double well,
$x,x'$ and $\vec{r} = (y,z)$ respectively denote longitudinal and
transverse coordinates, and $G(x,t)$ is the Green's function for a
free particle
\begin{align}
G(x,t) = \sqrt{\frac{m}{2 \pi i t \gamma}} \exp \left( i \frac{m}{2t \gamma} x^{2} \right). \label{eq:free_GF}
\end{align}
The function $f(\vec{r},t)$ is an overall envelope whose precise from
follows from the details of the trapping potential. By measuring
$\varrho_{\mathrm{tof}}$, the system collapses to a simultaneous
eigenstate of all $e^{\frac{i}{2}\phi_{a,s}(x,t)}$. The outcome of such measurements can be simulated if one has access to
distribution functions of the corresponding eigenvalues
$e^{\frac{i}{2}\varphi_{a,s}(x,t)}$. Such distribution functions will be computed in 
Sec. \ref{sub:full_distribution_functions}. In principle, the
observable (\ref{eq:density_tof_bos_longitudinal_approx}) also
contains small contributions from the density fields $\Pi_{a,s}(x)$
\cite{Nieuwkerk2018}. In order to treat these, the above description
of a projective measurement has to be preceded by a diagonalization of
the full observable, which now contains noncommuting fields. We do not
pursue this further here because these effects are expected to be small
in the regime where our low-energy approximation applies.

Experiments typically report results related to the quantity
\begin{align}
R(x_{0},\vec{r},t_{1},t_0) &=
\int_{x_{0}-\ell}^{x_{0}+\ell} dx \,
\varrho_{\mathrm{tof}}(x,\vec{r},t_{1},t_0)\nn
&=\rho_0\Big| f(\vec{r},t_{1})\Big|^2\int^{x_0+\ell}_{x_0-\ell}dx
\left[|g_+(x)|^2+|g_-(x)|^2+2{\rm
    Re}\Big(g_+(x)g_-^*(x)e^{i\frac{m\vec{r}\cdot\vec{d}}{t_1}} \Big)\right]
, \label{eq:density_tof_bos_longitudinal_approx_integrated} 
\end{align}
where we have defined
\be
g_\pm(x)=\int dx^{\prime}\, G(x-x^{\prime},t_{1}) e^{\frac{i}{2}\left(
  \varphi_{s}(x^{\prime},t_0)\pm \varphi_{a}(x^{\prime},t_0)   \right)}\ .
\label{gpm}
\ee

\subsection{Mode expansions for the two-component Luttinger liquid} 
\label{sub:mode_expansions_for_H0}

The free boson Hamiltonians $H_{a,s}$ are diagonalized by the mode
expansions (see e.g. \cite{Cazalilla2004})
\begin{align}
\theta_{j}(x) &= \theta_{j,0} + \frac{\pi x}{L} \delta N_{j} + i \sum_{q>0} \left(  \frac{\pi K}{qL} \right)^{1/2} \sin{q x} \left( b_{j,q}^{\vphantom{\dagger}} -b^{\dagger}_{j,q} \right), \label{eq:mass_quench_theta_modes} \\
\phi_{j}(x) &= \phi_{j,0} + \sum_{q>0} \left( \frac{\pi}{qKL} \right)^{1/2} \cos{q x} \left( b_{j,q}^{\vphantom{\dagger}} +b^{\dagger}_{j,q} \right),
\end{align}
where $q = \frac{\pi n}{L}$, $n \in \mathbb{Z}$, $\left[
  b^{\vphantom{\dagger}}_{q},b^{\dagger}_{k} \right] =
\delta^{\vphantom{\dagger}}_{q,k}$ and $\left[ \delta N_{j},
  \phi_{l,0} \right] = i \delta_{j,l} $. The zero modes $\delta N_{j}$
have integer eigenvalues. The Hamiltonians then take the form 
\begin{align}
H_{j} = \frac{v \pi}{2 L K}\delta N^{2}_{j}+ \sum_{q>0} v q \,b_{j,q}^{\dagger} b_{j,q}^{\vphantom{\dagger}}, \;\;\; j=a,s. \label{eq:ham_modes}
\end{align}
Going back to Eq. (\ref{eq:bosonization_identity}), we see that the hard-wall condition (\ref{eq:hard_wall_cond}) is guaranteed by choosing the c-number $\theta_{0}$ such that
\begin{align}
\theta(0) = \theta_{0} \notin \mathbb{Z}.
\end{align}
It turns out to be useful in what follows to rewrite the
mode-expansions in the form
\begin{align}
\phi_{l}(x,t) &=\sum_{\nu}u^{(l)}_{\nu}(x) \left( b_{\nu}^{\phdag}(t)
+ b_{\nu}^{\dagger}(t) \right),\label{eq:phi_vec_form_full} \\
\partial_{x} \theta_{l}(x,t)/\pi &= \sum_{\nu}w^{(l)}_{\nu}(x) \left(
b_{\nu}^{\phdag}(t) - b_{\nu}^{\dagger}(t) \right)\ ,\quad l=a,s\ . \label{eq:Pi_vec_form_full}
\end{align}
Here we have introduced a multi-index $\nu=(l,q)$ that runs over all positive momenta
$q\geq 0$ and the two sectors $l=a,s$ and we have defined
\begin{align}
u^{(l)}_{(j,q)}(x) &= \delta_{j,l}\begin{cases}
    \left( \frac{\pi}{q K L} \right)^{1/2} \cos qx , &\text{ if }q \neq 0\ ,\\
    \frac{1}{2}\sqrt{\frac{1}{K}}&\text{ if }q = 0\ ,
\end{cases} \label{eq:u_vect}\\
w^{(l)}_{(j,q)}(x) &= \delta_{j,l}\begin{cases}
    i \left( \frac{q K}{\pi L} \right)^{1/2} \cos qx, &\text{ if }q \neq 0\ ,\\
    \frac{i}{L} \sqrt{K}&\text{ if }q = 0\ ,
\end{cases}\\
b_{j,0} &= \sqrt{K} \phi_{j,0} - \frac{i}{2} \sqrt{\frac{1}{K}}\delta N_{j}\ .
\end{align}



\section{Self-consistent time-dependent harmonic approxiation} 
\label{sec:self_consistent_time_dependent_harmonia_approxiation}



Our aim is to determine the non-equilibrium evolution after a \emph{quantum
quench}: the system is prepared in a density matrix $\rho(0)$ that does not
commute with the Hamiltonian (\ref{eq:full_hamiltonian}). We moreover
take the density matrix to be Gaussian for simplicity. The ensuing
time evolution is described in the Schr\"odinger picture via the
time evolving density matrix  
\begin{align}
\rho(t) = e^{-iH t} \rho(0) e^{iH t}.
\label{eq:time evolved_density_matrix}
\end{align}
As our Hamiltonian of interest \fr{eq:full_hamiltonian} is not solvable
we resort to an analysis by means of a SCTDHA
\cite{Boyanovsky1998,Sotiriadis2010,Nieuwkerk2018b,Lerose19,Collura20}. Below we 
generalize the analysis of \cite{Nieuwkerk2018b} to include the
nonlinear interaction between the symmetric and antisymmetric sectors.
The SCTDHA amounts to  
replacing the exact time evolution operator with
\begin{align}
e^{-iH t}\longrightarrow U_{\rm SCH}(t)=Te^{-i \int_{0}^{t}
  H_{\mathrm{SCH}}(\tau) d \tau}\ , \label{eq:repl_time_evol}
\end{align}
where
\begin{align}
  H_{\mathrm{SCH}}(t)= H_a &+ H_s + \int dx \bigg[
    f(x,t) + \phi_{a}(x) g^{(1)}(x,t) \notag\\
&+ \Pi_{s}(x) g^{(2)}(x,t) + \phi_{a}^{2}(x) h^{(1)}(x,t) + \phi_{a}(x)\Pi_{s}(x) h^{(2)}(x,t)\bigg].
\label{eq:replacement_generic}
\end{align}
Here the functions $g^{(1,2)}(x,t)$ and $h^{(1,2)}(x,t)$ are
determined self-consistently. In order to derive
\fr{eq:replacement_generic} we decompose the fields into their space
and time dependent expectation values and their fluctuations
\begin{align}
\phi_{l}(x,t) &= \braket{\phi_{l}(x,t)} + \chi_{l}(x,t),\\
\Pi_{l}(x,t) &= \braket{\Pi_{l}(x,t)} + \pi_{l}(x,t)\ ,\quad l=a,s.
\end{align}
Substituting this decomposition into the interaction part of the Hamiltonian
(\ref{eq:mixing_hamiltonian}) gives
\begin{align}
H_{\perp} = - 2t_{\perp} \int_{0}^{L} dx \, \left[ \rho_{0} + \sigma
  \braket{\Pi_{s}} + \sigma \pi_{s} \right] \left[ \cos
  \braket{\phi_{a}} \cos \chi_{a} - \sin \braket{\phi_{a}} \sin
  \chi_{a} \right] . 
\end{align}
In the next step we expand the Hamiltonian to quadratic order in
fluctuations following \cite{Nieuwkerk2018b}, which gives
\begin{align}
H_{\perp}\approx - 2t_{\perp} \int dx \, \Bigg[ &\left(\rho_{0} + \sigma \,\pi_{s} - \frac{1}{2} \left( \rho_{0} + \sigma \,\left< \Pi_{s} \right> \right) \chi_{a}^{2}   - \sigma \left< \chi_{a} \,\pi_{s} \right> \chi_{a} \right) \cos \left< \phi_{a} \right> \\
- &\left( \left( \rho_{0} + \sigma \left( \pi_{s} + \braket{\Pi_{s}}
\right)  \right) \chi_{a} - \frac{\sigma}{2} \left< \chi_{a} \,\pi_{s}
\right> \chi_{a}^{2}  \right) \sin \left< \phi_{a} \right> \Bigg]
e^{-\frac{1}{2} \left< \chi_{a}^{2} \right> } + {\rm const} \notag
\end{align}
After re-expressing this in terms of the original fields $\phi_{a}$
and $\Pi_{s}$, we arrive at Eq. (\ref{eq:replacement_generic}), where
the functions $h^{(j)}(x,t)$ and $g^{(j)}(x,t)$ are determined
self-consistently by
\begin{align}
h^{(1)}(x,t) &= \mathrm{Re} \overline{F}(x,t)/2, \nn
h^{(2)}(x,t) &= \sigma \mathrm{Im} F(x,t),\nn
g^{(1)}(x,t) &= \mathrm{Im} \overline{F}(x,t) - 2 \braket{\phi_{a}(x,t)} h^{(1)}(x,t) - \braket{\Pi_{s}(x,t)} h^{(2)}(x,t), \nn
g^{(2)}(x,t) &= -\sigma \mathrm{Re} F(x,t) - \braket{\phi_{a}(x,t)} h^{(2)}(x,t). \label{eq:h_g_def4}
\end{align}
Here we have defined two functions
\begin{align}
F(x,t)&= 2t_{\perp} {\rm Tr}\bigg[U_{\rm SCH}(t)\rho(0)U^\dagger_{\rm SCH}(t)
  e^{i\phi_{a}(x)}\bigg], \nn
\overline{F}(x,t) &= 2t_{\perp}{\rm Tr}\bigg[U_{\rm
    SCH}(t)\rho(0)U^\dagger_{\rm SCH}(t) e^{i\phi_{a}(x)}
  \left(\rho_{0} + \sigma \Pi_{s}(x) \right)\bigg]. \label{eq:def_Fs}  
\end{align}
One subtlety associated with the SCTDHA concerns the zero mode
$\phi_{a,0}$. The spectrum of $\phi_{a,0}$ originally reflected the
compact nature of the phase field $\phi_a(x)=\phi_a(x)+2\pi$. The
latter feature is lost in the SCTDHA, where fluctuations are assumed
to be small but the fields themselves take arbitrary real values.


\subsection{Gaussian initial states} 
\label{sub:initial_state}


In order to investigate the effects of the $\sigma$-term that couples
the symmetric and antisymmetric sectors we want to start from a
factorized state and study how correlations develop over time. An important
requirement is related to our use of the SCTDHA: its accuracy strongly
depends on the initial state obeying Wick's theorem. These two
considerations lead us to consider the same class of initial states previously
used in the literature \cite{Bistritzer2007,Kitagawa2010,Kitagawa2011}
\begin{align}
\rho(0) = \rho_a(0)\otimes \rho_{s}(0)\ ,
\end{align}
where $\rho_a(0)=|V, r,  \varphi\rangle_{a}{}_a\langle V,r,\varphi|$ is a Gaussian pure state 
\begin{align}
\ket{V, r, \varphi}_{a} = \mathcal{N} \exp \left(\sum_{pq} V_p \left( \sech r^T
\right)_{pq} b^{\dagger}_{a,q} +\sum_{p,q,k} \frac{1}{2} b^{\dagger}_{a,p} \left( \tanh r
\right)_{pq}  e^{i \varphi_{q,k}} b^{\dagger}_{a,k} \right)|0\rangle_a
. \label{eq:general_gaussian} 
\end{align}
It is useful to define new annihilation operators $\alpha_{a,k}$ satisfying
\begin{align}
\alpha_{a,k}\ket{V, r, \varphi}_{a}=0\ ,
\end{align}
which are related to the $b$-operators via the canonical transformation
\begin{align}
b_{a,q} = \sum_{k}( \cosh r )_{qk}  \left[ \alpha_{a,k} + V_{k} \right] + \left(\sinh r e^{i \varphi}\right)_{qk}\left[ \alpha^{\dagger}_{a,k} + V^{*}_{k} \right]. \label{eq:a_as_alphas}
\end{align}
In previous works it has been assumed that the symmetric
sector is initialized in a thermal state \cite{Kitagawa2011}. We
will follow this assumption, but in order to study the effects of spatial
inhomogeneity we take our initial state to be given by a ``displaced''
thermal density matrix
\begin{align}
\rho_{s} = D(R)\frac{e^{- \beta H_{s}}}{\mathrm{Tr}\,e^{- \beta H_{s}}} D^{\dagger}(R)\ ,
\end{align}
where the displacement operators are defined via
\begin{align}
D^{\dagger}(R) b_{j,k} D(R) &= b_{j,k} + R_{j,k}\ , \quad j=a,s \ .
\end{align}
This suggests the definition of displaced annihilation operators $\alpha_{s,k}$ via a constant shift
\begin{align}
b_{s,k} = \alpha_{s,k} + R_{s,k}\ , \label{eq:def_alpha_s}
\end{align}
so that
\begin{align}
\braket{\alpha_{s,k}} = 0
\end{align}
on the initial state. Since $\rho_{s}(0)$ satisfies Wick's theorem, it is then completely fixed by the vector $R_{s,k}$ along with connected two-point functions of the fields. Using the mode expansion of $H_{s}$ from Eq. (\ref{eq:ham_modes}) we simply find bosonic occupation numbers for $q>0$, 
\begin{align}
\left< b^{\dagger}_{s,q} b^{\phdag}_{s,k} \right>_{c}  = \frac{\delta_{q,k}}{e^{\beta v q}-1} \equiv n_{(s,q)},
\end{align}
the anomalous expectation values $\braket{b_{s,q} b_{s,q^{\prime}}}_{c}$ being zero. For the zero mode, the only expectation values on $\rho_{s}(0)$ that we will need are
\begin{align}
\braket{\delta N_{s}^{2}}_{c} = \frac{\sum_{n} e^{- \beta \frac{v \pi}{2 K L} n^{2}} n^{2}}{\sum_{n} e^{- \beta \frac{v \pi}{2 K L} n^{2}}}, \quad \braket{\delta N_{s}} = 0,
\end{align}
where the second identity implies $\mathrm{Im}R_{s,0}(0)=0$. As will
become clear in the next section, expectation values involving the
field $\phi_{s,0}$ will never be required for the computation of
physical quantities. 

\subsection{Equations of motion} 
\label{sub:equations_of_motion}
The SCTDHA allows for a closed-form expression of the equations of
motion. We will work in the Heisenberg picture from here onwards. The
SCTDHA guarantees that time evolving annihilation operators can always
be written as 
\begin{align}
b_{\nu}(t) = R_{\nu}(t) + S_{\nu \mu}^{\phdag}(t)
\alpha_{\mu}^{\phdag} + T_{\nu \mu}^{*}(t)
\alpha_{\mu}^{\dagger} \label{eq:annihilation_operator_generic} 
\end{align}
where $\alpha_{\mu}$ are a set of bosonic creation and annihilation
operators. We choose these to be given by
\be
\alpha_{\nu}=\begin{cases}
\alpha_{a,k} & \text{if } \nu=(a,k)\\
\alpha_{s,k} & \text{if } \nu=(s,k)
\end{cases},
\ee
where the $\alpha_{a,k}$ are defined in \fr{eq:a_as_alphas} and the
$\alpha_{s,k}$ in (\ref{eq:def_alpha_s}). For \fr{eq:annihilation_operator_generic} to be a canonical
transformation we require  
\begin{align}
    S S^{\dagger} - T^{*} T^{T} = \mathbbm{1}\ ,\qquad
    S T^{\dagger} - T^{*} S^{T} = 0\ . \label{eq:Canonical_Condition_matrix}
\end{align}
The initial condition on $R, S$ and $T$ are given by
\begin{align}
R_{\mu}(0) &= \begin{cases}
\sum_{q}(\cosh r)_{pq} \, V_q + (\sinh r e^{i \varphi})_{pq}
\, V^{*}_q& \text{if }\mu=(a,p),\\
0 & \text{else,}
\end{cases} \notag\\
S_{\nu,\mu}(0) &= \begin{cases}
( \cosh r )_{pq} & \text{if } \nu=(a,p),\ \mu=(a,q),\\
\delta_{pq}& \text{if } \nu=(s,p),\ \mu=(s,q),\\
0 & \text{else, }
\end{cases}
\label{eq:IC_RST}\\
T^*_{\nu,\mu}(0) &= \begin{cases}
( \sinh r e^{i\varphi})_{pq} & \text{if } \nu=(a,p),\ \mu=(a,q),\\
0 & \text{else.}
\end{cases}\notag
\end{align}
We note that the $\alpha_{\mu}$'s satisfy Wick's theorem on the initial state,
along with $\left< \alpha_{\mu} \right> =0$ for all $\mu$. 

The time evolution of any operator is then encoded in the time-dependence of the tensors $R, S$ and $T$, which we will now determine. To this end, we write the SCTDHA Hamiltonian in the generic form
\begin{align}
H_{\mathrm{SCH}}(t) = b^{\dagger}_{\nu} A_{\nu\mu}^{\phdag}(t) b^{\vphantom{\dagger}}_{\mu} &+ \frac{1}{2} \left( b_{\nu}^{\dagger} B_{\nu\mu}^{\dagger}(t) b_{\mu}^{\dagger} + b_{\nu} B_{\nu\mu}(t) b_{\mu} \right) \notag\\
&+ C(t)+ D^{\vphantom{\dagger}}_{\nu}(t) \left( b^{\vphantom{\dagger}}_{\nu} + b_{\nu}^{\dagger} \right) + E^{\vphantom{\dagger}}_{\nu}(t) \left( b^{\vphantom{\dagger}}_{\nu} - b_{\nu}^{\dagger} \right). \label{eq:generia_quadratic_ham}
\end{align}
The matrices $A,B$ and vectors $D,E$ depend on the self-consistency
functions $g^{(1,2)}$ and $h^{(1,2)}$, cf. Eqs. (\ref{eq:h_g_def4}),
and are given in Appendix
\ref{sec:tensors_occurring_in_HSCH}. Inserting the expansion
(\ref{eq:annihilation_operator_generic}) into the Heisenberg equation
of motion 
\begin{align}
i \frac{d}{dt} b_\nu(t) = U_{\rm SCH}(t)\left[ b_\nu, H_{\mathrm{SCH}}(t)\right]U^\dagger_{\rm SCH}(t) \label{eq:Heisenberg_EOM}
\end{align}
yields a system of coupled, first order differential equations
\begin{align}
i\dot{R}_{\nu}(t) &= A_{\nu\mu}(t)R_{\mu}(t) + B^{\dagger}_{\nu\mu}(t) R_{\mu}^{*}(t) + D_{\nu}(t) - E_{\nu}(t) \notag\\
i\dot{S}_{\nu\mu}(t) &= A_{\nu\lambda}(t)S_{\lambda\mu}(t) + B^{\dagger}_{\nu\lambda}(t) T_{\lambda\mu}(t) \label{eq:ODEs}\\
-i\dot{T}_{\nu\mu}(t) &= A^{*}_{\nu\lambda}(t)T_{\lambda\mu}(t) + B^{T}_{\nu\lambda}(t) S_{\lambda\mu}(t). \notag
\end{align}
This system of ODE's is \textit{nonlinear}: as a result of the
self-consistency functions (\ref{eq:h_g_def4}) on
which the tensors $A,B,D$ and $E$ depend, these tensors are themselves
functions of $R, S$ and $T$, which therefore enter the system
(\ref{eq:ODEs}) in nonlinear combinations.  
To simplify some of the following equations we introduce linear combinations
\begin{align}
Q^{\phdag}_{\nu \mu}(t) = S^{\phdag}_{\nu \mu}(t) + T^{\phdag}_{\nu \mu}(t), \;\;\;\;
\overline{Q}^{\phdag}_{\nu \mu}(t) = S^{\phdag}_{\nu \mu}(t) - T^{\phdag}_{\nu \mu}(t)\ . \label{eq:def_Q} 
\end{align}
In terms of these functions mode expansions of the time evolved fields
take the form
\begin{align}
    \phi_{a}(x,t) &= \sum_{\nu} u^{(a)}_{\nu}(x) \left( 2 \mathrm{Re}
    R^{\vphantom{*}}_{\nu}(t) + \sum_{\mu} \left[
      Q^{\phdag}_{\nu\mu}(t) \alpha_{\mu}^{\vphantom{\dagger}} +
      Q^{*}_{\nu \mu}(t) \alpha_{\mu}^{\dagger} \right]
    \right) \label{eq:phi_as_alphas}\ ,\\
    \Pi_{l}(x,t) &= \sum_{\nu} w^{(l)}_{\nu}(x) \left( 2i \mathrm{Im} R^{\vphantom{*}}_{\nu}(t) + \sum_{\mu}\left[ \overline{Q}_{\nu \mu}(t) \alpha_{\mu}^{\vphantom{\dagger}} - \overline{Q}^{*}_{\nu \mu}(t) \alpha_{\mu}^{\dagger} \right]  \right). \label{eq:Pi_as_alphas}
\end{align}
The functions (\ref{eq:def_Fs}) can then be computed using Wick's
theorem for the $\alpha$-operators, based on the above
expressions. This closes the system of ODE's (\ref{eq:ODEs}). The zero
mode in the symmetric sector $\phi_{s,0}$ reflects the compact nature
of the phase field $\phi_s$ and therefore needs to be treated
separately from the finite momentum modes. We therefore define a field
\begin{align}
\widetilde{\phi_{s}}(x) \equiv \phi_{s}(x) - \phi_{s,0}, 
\end{align}
which time evolves as
\begin{align}
\widetilde{\phi_{s}}(x,t) &= \sum_{\nu\neq (s,0)} u^{(s)}_{\nu}(x) \left( 2 \mathrm{Re} R^{\vphantom{*}}_{\nu}(t) + \sum_{\mu} \left[ Q^{\phdag}_{\nu\mu}(t) \alpha_{\mu}^{\vphantom{\dagger}} + Q^{*}_{\nu \mu}(t) \alpha_{\mu}^{\dagger} \right]  \right).
\end{align}
Importantly the zero mode $\phi_{s,0}$ does not get generated under
Heisenberg time evolution of other fields. This is easily checked by
inspection of the Hamiltonian (\ref{eq:full_hamiltonian}) which is
seen to not involve $\phi_{s,0}$. This in turn implies that the zero
mode cannot appear on the rhs of the Heisenberg equation of motion
(\ref{eq:Heisenberg_EOM}). Since we can express the zero mode at $t=0$ as 
\begin{align}
\phi_{s,0} = \left( \alpha_{(s,0)}^{\phdag} + \alpha_{(s,0)}^{\dagger} \right)/\sqrt{4K},
\end{align}
we conclude that this linear combination of $\alpha$-operators does
not appear in the sums over modes in
(\ref{eq:phi_as_alphas},\ref{eq:Pi_as_alphas}) except in the
expansion for $\phi_{s}(x,t)$, where it occurs in the term with $\nu = 
(s,0)$. This directly leads to 
\begin{align}
\mathrm{Re}\, Q_{\nu,(s,0)}(t) &= 0 \;\;\; \forall \;\nu \neq (s,0) ,\qquad
\mathrm{Im}\, \overline{Q}_{\nu,(s,0)}(t) = 0 \;\;\; \forall \; \nu.
\end{align}

\subsection{Self-consistent expectation values} 
\label{sub:self_consistent_expectation_values}

\subsubsection{One-point functions}

As all relevant one-point functions of $\alpha_{\nu}$ and $\delta
N_{s}$ are zero we have
\begin{align}
    \left< \widetilde{\phi_{s}}(x,t) \right>  &= 2\sum_{\nu \neq (s,0)}
    u^{(s)}_{\nu}(x) \mathrm{Re}
    R^{\vphantom{*}}_{\nu}(t) \label{eq:phi_s_1pt}\ , \\
    \left< \phi_{a}(x,t) \right>  &= 2\sum_{\nu} u^{(a)}_{\nu}(x)
    \mathrm{Re} R^{\vphantom{*}}_{\nu}(t) \label{eq:phi_a_1pt}\ ,\\
    \left< \Pi_{l}(x,t) \right>  &= 2i \sum_{\nu} w^{(l)}_{\nu}(x) \mathrm{Im} R^{\vphantom{*}}_{\nu}(t). \label{eq:Pi_1pt}
\end{align}
\subsubsection{Two-point functions}
Comparing the definitions from Section \ref{sub:initial_state} to the
initial conditions (\ref{eq:IC_RST}), we find that  for any $\nu, \mu
\neq (s,0)$, 
\begin{align}
{\mathfrak g}_{\nu,\mu}=\left< \alpha_{\nu}^{\dagger} \alpha_{\mu}^{\phdag} \right> &=
  \left< \alpha_{\nu}\alpha_{\mu}^{\dag} \right> -\delta_{\nu,\mu}=
\delta_{\nu,\mu}\begin{cases}
  0 & \text{if } \nu\in\{(a,q),(s,0)\}\\
  n_{(s,q)} & \text{if } \nu\in\{(s,q)|q\neq 0\}
\end{cases}.
\end{align}
If we define $P^{(s)}_{0}$ to be the projector on the symmetric zero modes, along with its complement $\tilde{\mathbbm{1}} = \mathbbm{1} - P^{(s)}_{0}$, we then find the following connected two-point functions
\begin{align}
\left< \phi_{j}(x,t)\phi_{l}(y,t) \right>_{c} &= u^{(j)}(x) \bigg(2{\rm Re}(Q^{*} \mathfrak{g} Q^{T})+ Q \tilde{\mathbbm{1}} Q^{\dagger}
+ \frac{\braket{\delta N^{2}_{s0}}}{K} \mathrm{Im}Q P^{(s)}_{0} \mathrm{Im}Q^{T} \bigg) u^{(l)}(y)\ , \label{eq:phi_2pt} \nn
\left< \phi_{j}(x,t)\Pi_{l}(y,t) \right>_{c} &= -u^{(j)}(x) \bigg(2i{\rm Im}(Q \mathfrak{g}\overline{Q}^{\dagger})+
Q \tilde{\mathbbm{1}} \overline{Q}^{\dagger}
+i \frac{\braket{\delta N^{2}_{s0}}}{K} \mathrm{Im}Q P^{(s)}_{0} \mathrm{Re}\overline{Q}^{T} \bigg) w^{(l)}(y)\ .
\end{align}
In the above, indices on all matrices and vectors have been suppressed for conciseness. If we want to consider the field $\widetilde{\phi_{s}}$ instead of $\phi_{s}$, we need leave out the symmetric zero mode term. This leads, for instance, to
\begin{align}
\left< \widetilde{\phi_{s}}(x,t)\Pi_{l}(y,t) \right>_{c} &= u^{(j)}(x)
\left(P^{(s)}_{0}-\mathbbm{1} \right) \times \notag \\
&\times\bigg(2i{\rm Im}(Q \mathfrak{g}\overline{Q}^{\dagger})+
Q \tilde{\mathbbm{1}} \overline{Q}^{\dagger}
+i \frac{\braket{\delta N^{2}_{s0}}}{K} \mathrm{Im}Q P^{(s)}_{0} \mathrm{Re}\overline{Q}^{T} \bigg)w^{(l)}(y)\ ,
\end{align}
and analogous modifications for $\left< \widetilde{\phi_{s}}(x,t)\widetilde{\phi_{s}}(y,t) \right>_{c}$ and $\left< \widetilde{\phi_{s}}(x,t)\phi_{a}(y,t) \right>_{c}$.

\subsection{Full distribution functions} 
\label{sub:full_distribution_functions}

Individual measurement outcomes in interference experiments of
interest \cite{Pigneur2017} are fully determined by the eigenvalues
$\varphi_{a}$ and $\widetilde{\varphi_{s}}$ of the phase fields
$\phi_{a}$ and $\widetilde{\phi_{s}}$ \cite{Nieuwkerk2018},
cf. Eq. (\ref{eq:density_tof_bos_longitudinal_approx}). To model
the outcomes of such measurements we therefore require the
time-dependent distribution functions for $\varphi_{a}$ and
$\widetilde{\varphi_{s}}$. These can be determined in the framework of
the SCTDHA \cite{Nieuwkerk2018b,Collura20}. For the case at hand, we first expand the eigenvalues of the phase
fields as Fourier series,  
\begin{align}
\widetilde{\varphi_{s}}(x,t) &= \sum_{\mu \neq (s,0)} u^{(s)}_{\mu}(x)
f_{\mu,t}\ ,\qquad
\varphi_{a}(x,t) = \sum_{\mu} u^{(a)}_{\mu}(x)f_{\mu,t}\ . \label{eq:phase_evals}
\end{align}
Here we have again used our multi-index notations $\mu = (j,q)$, where
$j=a,s$ labels the sector and $q$ the momentum. Each measurement
selects a particular set of Fourier coefficients and we denote the
averages over many measurements by
\be
\overline{ f_{\mu,t}}\ ,\quad
\overline{ f_{\mu,t}\ f_{\nu,t}}\quad \text{etc}.
\ee
The mean values for the Fourier coefficients can be read off from the
one-point functions calculated earlier, \emph{cf.}
Eqs. (\ref{eq:phi_s_1pt},\ref{eq:phi_a_1pt}) 
\begin{align}
\overline{{f}_{\mu,t}} = 2\mathrm{Re}\, R_{\mu}(t)\ . \label{eq:fs_mean}
\end{align}
The object of interest is then the time-dependent joint probability
distribution $P$ of Fourier coefficients $\{\mathfrak{f}_{\mu} \}$. Within
the SCTDHA all cumulants of $\phi_{a,s}$ other than the variance
vanish, so that this  probability distribution is Gaussian
\begin{align}
P(\{\mathfrak{f}_{\mu} \},t) = \frac{1}{(2 \pi)^{N/2}}
\frac{1}{\sqrt{\det M(t)}} \mathrm{exp} \left( - \frac{1}{2}
\sum_{\mu,\nu}\left( \mathfrak{f}_{\mu} - \overline{f_{\mu,t}} \right)
M^{-1}_{\mu \nu}(t)\left( \mathfrak{f}_{\nu} - \overline{f_{\nu,t}}
\right)  \right)\ . \label{eq:prob_distr_fs} 
\end{align}
Here $N$ is the total number of Fourier modes retained in
\fr{eq:phase_evals}. Noting that
\begin{align}
  \left< \phi_{j}(x,t)\phi_{l}(y,t) \right>_{c} = u^{(j)}_{\mu}(x)
 \big( \overline{f_{\mu,t}f_{\nu,t}}-\overline{f_{\mu,t}}\ \overline{f_{\nu,t}}\big)
  u^{(l)}_{\nu}(y), \;\;\;\; j,l \in \{a,s\}
\end{align}
and comparing to Eq. (\ref{eq:phi_2pt}), we can directly read off the covariance matrix as well:
\begin{align}
M(t) = 2{\rm Re}(Q \mathfrak{g} Q^{\dagger}) + QQ^{\dagger} + \frac{\braket{\delta N^{2}_{s0}}}{K} \mathrm{Im}Q P^{(s)}_{0} \mathrm{Im}Q^{T}. \label{eq:fs_cov_mat}
\end{align}
Having obtained a time-dependent probability distribution for the
coefficients $\{f_{\mu,t} \}$, we can directly model experiments: we
draw coefficients $\{f_{\mu,t} \}$ from the distribution
(\ref{eq:prob_distr_fs}), reconstruct the corresponding eigenvalues
(\ref{eq:phase_evals}), and insert these in the time-of-flight density
(\ref{eq:density_tof_bos_longitudinal_approx}) to compute the measured
density profile. We note that in the notations used above the set
$\{\mathfrak{f}_{\mu} \}$ contains the non-physical Fourier coefficient
$\mathfrak{f}_{(s,0)}$. This quantity does not enter the observable
(\ref{eq:density_tof_bos_longitudinal_approx}), and can simply be
discarded, whenever a set of coefficients is drawn from $P \left(
\{\mathfrak{f}_{\mu} \}, t \right) $. 

By repeating the above procedure for modelling a measurement many
times over we can reconstruct the full distribution function of any
observable that depends only on the phase fields $\phi_{a,s}$. In what
follows, we will focus on the ``interference term'' in the spatially
integrated density after time-of-flight
$R_{\mathrm{tof}}(x_{0},\vec{r},t_{1},t_0)$ defined in
(\ref{eq:density_tof_bos_longitudinal_approx_integrated}). The  
eigenvalues of this observable are proportional to  
\begin{align}
\mathcal{I}_{\ell}\left( \{f_{\mu} \},x_{0},t_{0},t_{1} \right) =
\frac{1}{\ell} &\int_{x_{0}-\ell/2}^{x_{0}+\ell/2} dx\, g_+(x)g_-^*(x)
\label{eq:interference_term}
\end{align}
where $g_\pm(x)$ are defined in \fr{gpm} and are related to the
coefficients $f_{\mu}$ via (\ref{eq:phase_evals}).
Motivated by the experimental data analyses of
Refs~\cite{Kuhnert2013,Smith2013,Pigneur2018} we parametrize the 
interference term (\ref{eq:interference_term}) as 
\begin{align}
\mathcal{I}_{\ell}\left( \{f_{\mu} \},x_0,t_0,t_1 \right) =
C_\ell(x_0,t_0,t_1,\{f_{\mu} \}) e^{i \Phi_\ell(x_0,t_0,t_1,\{f_{\mu}
  \})}\ . \label{eq:interference_eval_parametr} 
\end{align}
By drawing many sets $\{\mathfrak{f}_{\mu} \}$ of coefficients from
the distribution function $P \left(\{\mathfrak{f}_{\mu} \}, t \right)$
and plotting the resulting values of $\Phi_{\ell}$ or $C_{\ell}$ in a
normalized histogram, we converge to probability distributions
$P_{\Phi_{\ell},C_{\ell}}$ for these quantities. These distribution
functions can formally be written as 
\begin{align}
P_{\Phi_{\ell}}(\alpha,t_{0},t_{1}) &= \left( \prod_{\mu}\int d \mathfrak{f}_{\mu} \right) \delta \left( \alpha - \mathrm{Arg} \, \mathcal{I}_{\ell}\left( \{f_{\mu} \},x_{0},t_{0},t_{1} \right) \right)  P \left(\{\mathfrak{f}_{\mu} \}, t_{0} \right)\ , \\
P_{C_{\ell}}(\gamma,t_{0},t_{1}) &= \left( \prod_{\mu}\int d \mathfrak{f}_{\mu} \right) \delta \left( \gamma - \mathrm{Abs} \, \mathcal{I}_{\ell}\left( \{f_{\mu} \},x_{0},t_{0},t_{1} \right) \right)  P \left(\{\mathfrak{f}_{\mu} \}, t_{0} \right)\ .
\end{align}


\section{Results for experimentally relevant initial states} 
\label{sec:results_for_experimentally_relevant_initial_states}

\subsection{Choice of initial state} 
\label{sub:specific_initial_state}

We now specialize to an initial state that is often used in the
literature, see e.g. \cite{Bistritzer2007,Kitagawa2010,Kitagawa2011}. In these works,
a quasi-classical argument is used to conjecture how the state of a
pair of elongated Bose gases follows from the splitting process of a
single gas. It is reasoned that when splitting a gas, each particle
has an equal probability to end up in well $1$ or in well $2$. The
relative particle number resulting from this poisson process is thus a
stochastic variable with mean zero and variance proportional to the
particle density. Assuming short-range correlations, one arrives at 
\begin{align}
\left< \Pi_{a}(x,0) \Pi_{a}(y,0) \right>_{c} = \frac{\eta \rho_{0}}{2} \delta_{\xi}(x-y), \label{eq:dens_two_pt_init_Kitagawa}
\end{align}
with $\eta$ a phenomenological parameter which we will set to $1$. Following \cite{Kitagawa2011}, the delta function above is understood as a flat sum over plane waves running up to momentum $\pi/\xi$. To reproduce this initial two-point function, it suffices to use the initial state (\ref{eq:general_gaussian}), with $r$ a real and diagonal matrix and $\varphi = 0$. The resulting initial condition on $\overline{Q}$,
\begin{align}
\overline{Q}(0)_{(a,j)(a,k)} = \delta_{jk} e^{-r_{jj}}\ ,
\end{align}
then leads to
\begin{align}
\left< \Pi_{a}(x,0)\Pi_{a}(y,0) \right>_{c} = \frac{K}{L^{2}} e^{-2r_{00}} + \sum_{j>0} \frac{qK}{\pi L} \cos \left( q_{j} x \right) \cos \left( q_{j} y \right) e^{-2 r_{jj}}\ . \label{eq:phi_tpt_generic_IC}
\end{align}
Comparing Eqs. (\ref{eq:dens_two_pt_init_Kitagawa}) and (\ref{eq:IC_RST}), we can thus read off
\begin{align}
e^{-2 r_{jk}} = \delta_{jk}\begin{cases}
    \frac{L \eta \rho_{0}}{2K} &\text{ if } q=0\ ,\\
    \frac{\pi \eta \rho_{0}}{q K} &\text{ if } q>0\ , 
\end{cases}
\end{align}
for the antisymmetric sector.

For the symmetric sector, we again follow Ref. \cite{Kitagawa2011}: the above quasiclassical splitting argument applies to the relative degrees of freedom, leaving the symmetric combinations of densities and phases unaltered. In \cite{Kitagawa2011}, the symmetric sector is therefore taken to be in a finite temperature equilibrium state. We adhere to this conjecture here and use the thermal density matrix described in Section \ref{sub:initial_state}, thereby fixing the initial conditions for both $T$ and $S$ in conjunction with the above discussion. Finally, the initial conditions for $R$ can be used to enforce various initial profiles on the density and phase fields in both sectors, which we will explore in Sec. \ref{sub:time_evolution} below.


\subsection{Experimental parameters} 
\label{sub:experimental_parameters}

We fix the parameters for our plots by following
Ref.~\cite{Kuhnert2013}: the one-dimensional density is taken to be
$\rho_{0} = 45 \, \mu \mathrm{m}^{-1}$, the healing length is $\xi=
\hbar\pi / mv = \pi \times 0.42 \,\mu \mathrm{m}$, the sound velocity
is given by $v \approx 1.738 \cdot 10^{-3} \, \mathrm{m}/\mathrm{s}$
and the Luttinger parameter in our conventions is $K \approx 28$. We
take the one-dimensional box size as large as we can achieve for a
given value of the cutoff length scale, which amounts to $L = 80 \,\mu
\mathrm{m}$. This is comparable to the size reported in
\cite{Kuhnert2013}. We work at a temperature of $5 \,\mathrm{nK}$
throughout. In all figures, time is measured in units of the
\textit{traversal time} \cite{EFreview}, $t_{\mathrm{tr}}=L/2v$, which
is the time it takes for a light cone to reach the edge of the system
from the centre of the box. We have chosen the value of the
phenomenological tunnel coupling strength $t_{\perp}$ by considering a
trade-off: we would like to maximize the Josephson frequency in order
to follow as many density-phase oscillations as possible, whilst
keeping the gap $\Delta$ of the model's dispersion relation no larger
than a small fraction of the energy cutoff in the Luttinger
liquid. The latter is equal to $\epsilon_{c} = v \pi / \xi$, with
$\xi$ the cutoff length scale. We have aimed for the ratio of the gap
to the cutoff to be no larger than $\Delta/\epsilon_{c} = 0.125$,
which we can guarantee by taking $t_{\perp} = 15 \,\mathrm{Hz}$ for the above
parameters. The only exception to the above is
Fig. \ref{fig:BC_compare}, where we take $t_{\perp} \approx
1.17\,\mathrm{Hz}$ following Ref. \cite{Nieuwkerk2018b}, to enable a
comparison with the case of periodic boundary conditions as presented
there.


\subsection{Time evolution} 
\label{sub:time_evolution}

We now consider time evolution under the SCTDHA Hamiltonian
(\ref{eq:replacement_generic}), with the initial condition described
in Sec. \ref{sub:specific_initial_state}. Throughout, we choose
$R(0)$ such that 
\begin{align}
\left< \phi_{a}(x,0) \right> = 0.2, \;\;\;\;\left< \Pi_{a}(x,0) \right> = 0. \label{eq:profiles_flat_a}
\end{align}
The one-point functions $\big< \widetilde{\phi_{s}}(x,0) \big>$ and
$\left< \Pi_{s}(x,0) \right>$ will be given different spatial
profiles, to investigate the effects of broken translational invariance. 
\subsubsection{\secfix{No coupling between symmetric and antisymmetric sectors
($\sigma=0$)}}
\begin{figure}[htbp!]
	\centering
	(a)\includegraphics[width=0.46\textwidth]{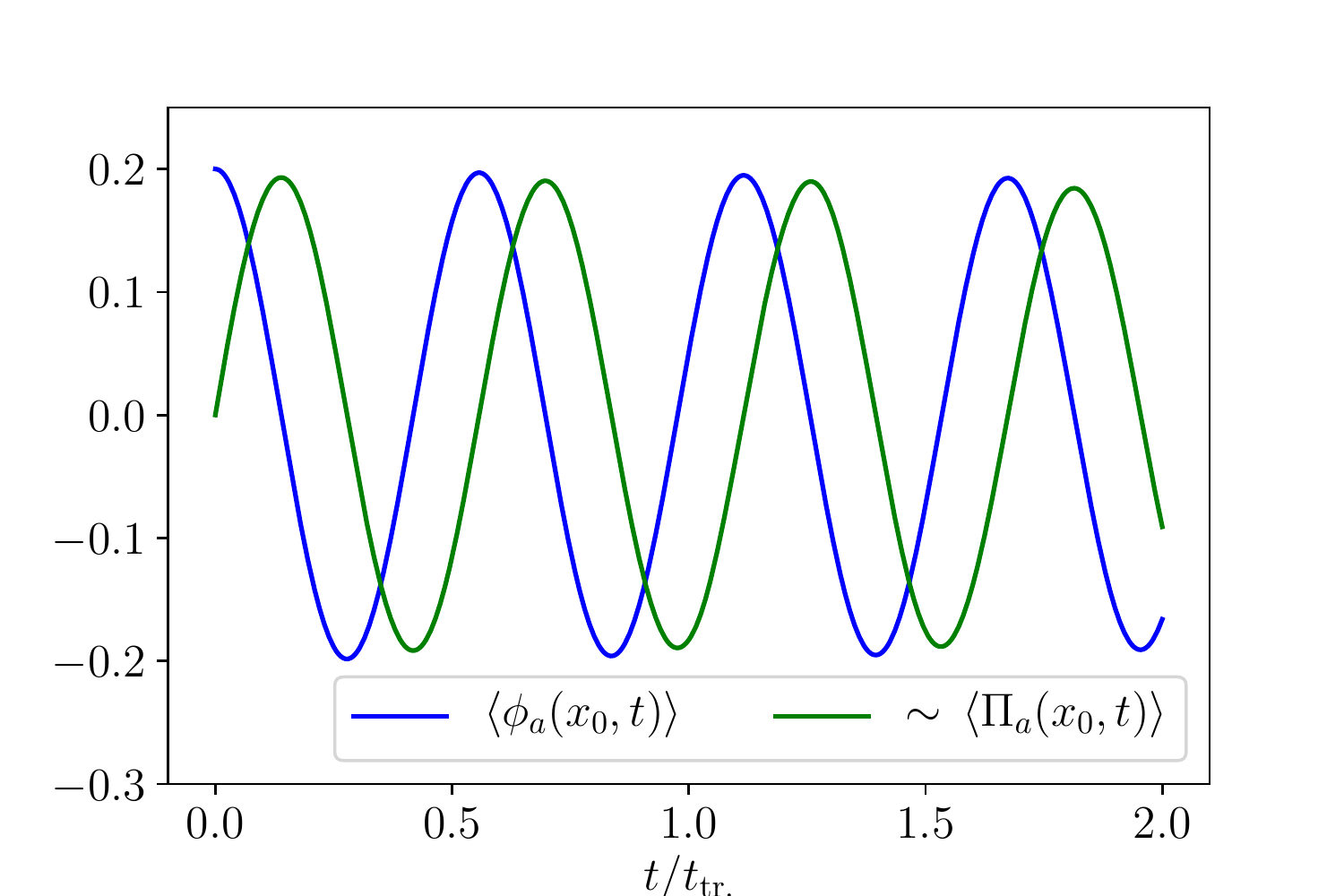}
	(b)\includegraphics[width=0.46\textwidth]{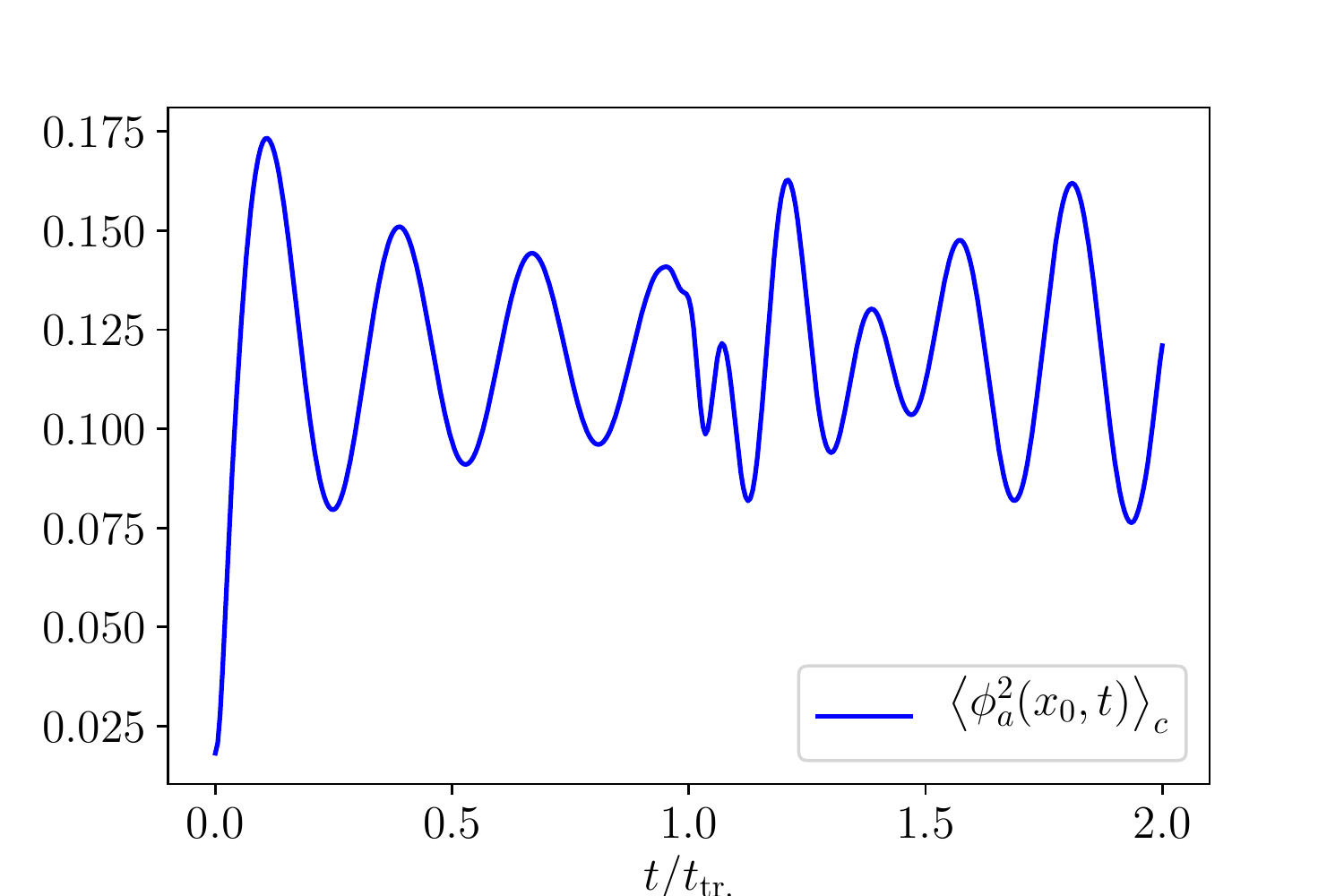}
	\caption{(a) Josephson oscillations of relative density (arbitrary units) and phase (radians) at the centre of the gas, $x_{0} = L/2$. (b) initial growth and oscillations of the variance of the relative phase. The initial phase and density profiles profiles are chosen according to Eqs. (\ref{eq:profiles_flat_a},\ref{eq:profiles_flat}) and coupling between the sectors is absent in these pictures, meaning $\sigma = 0$.}
	\label{fig:benchmark_plots}
\end{figure}

We will start from the situation where
\begin{align}
 \left< \widetilde{\phi_{s}}(x,0) \right> = 0 = \left< \Pi_{s}(x,0) \right>\ . \label{eq:profiles_flat}
 \end{align}
and $\sigma = 0$. This will serve as our benchmark, as it most closely
resembles the translationally invariant scenario described in
\cite{Nieuwkerk2018b} in which the (anti)symmetric sectors remain
uncorrelated. It is characterized by Josephson oscillations between
density and phase, see Fig. \ref{fig:benchmark_plots}(a), with a phase
variance that initially grows, and then shows oscillating behavior,
see Fig. \ref{fig:benchmark_plots}(b).

\begin{figure}[htbp!]
	\centering
	(a)\includegraphics[width=0.46\textwidth]{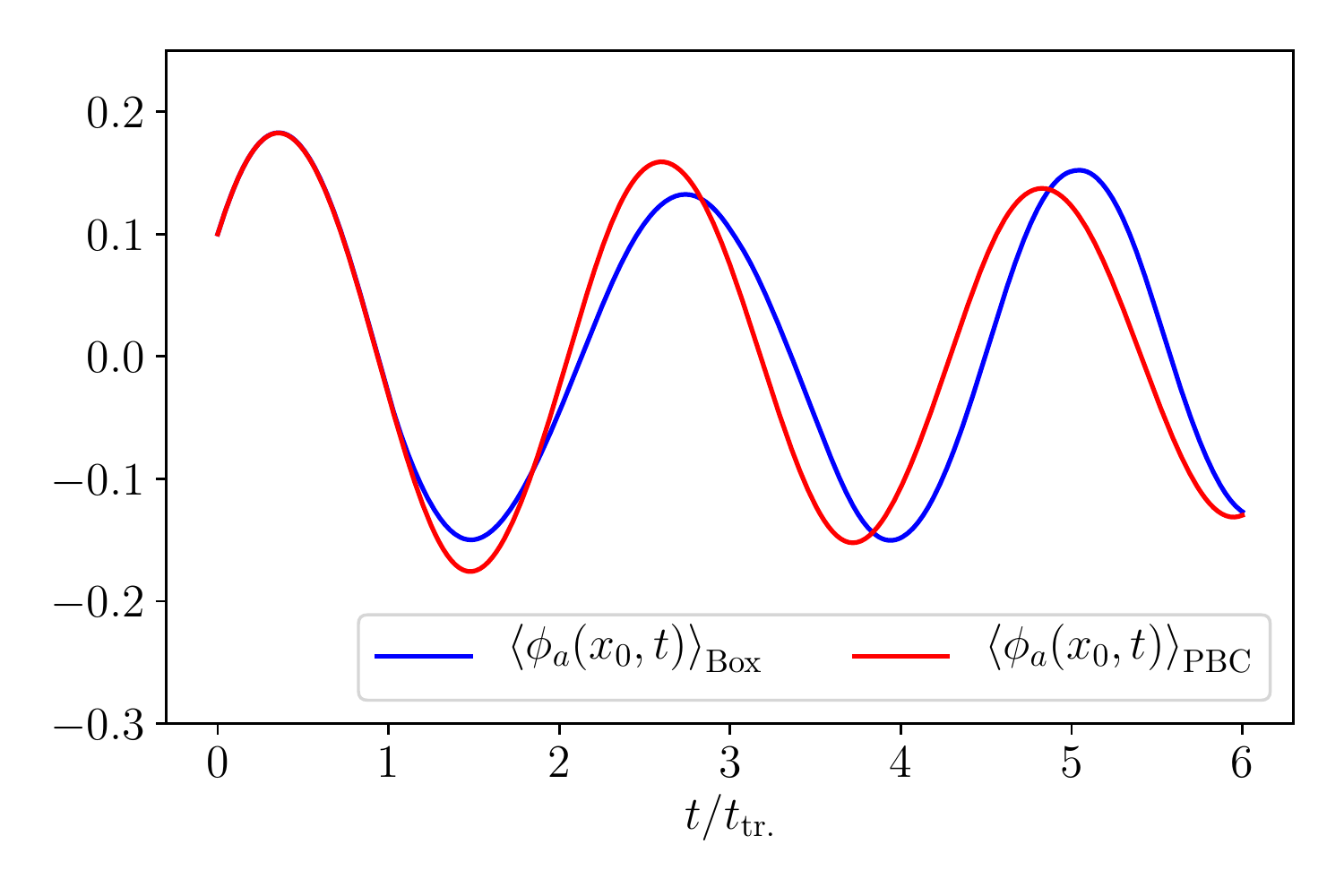}
	(b)\includegraphics[width=0.46\textwidth]{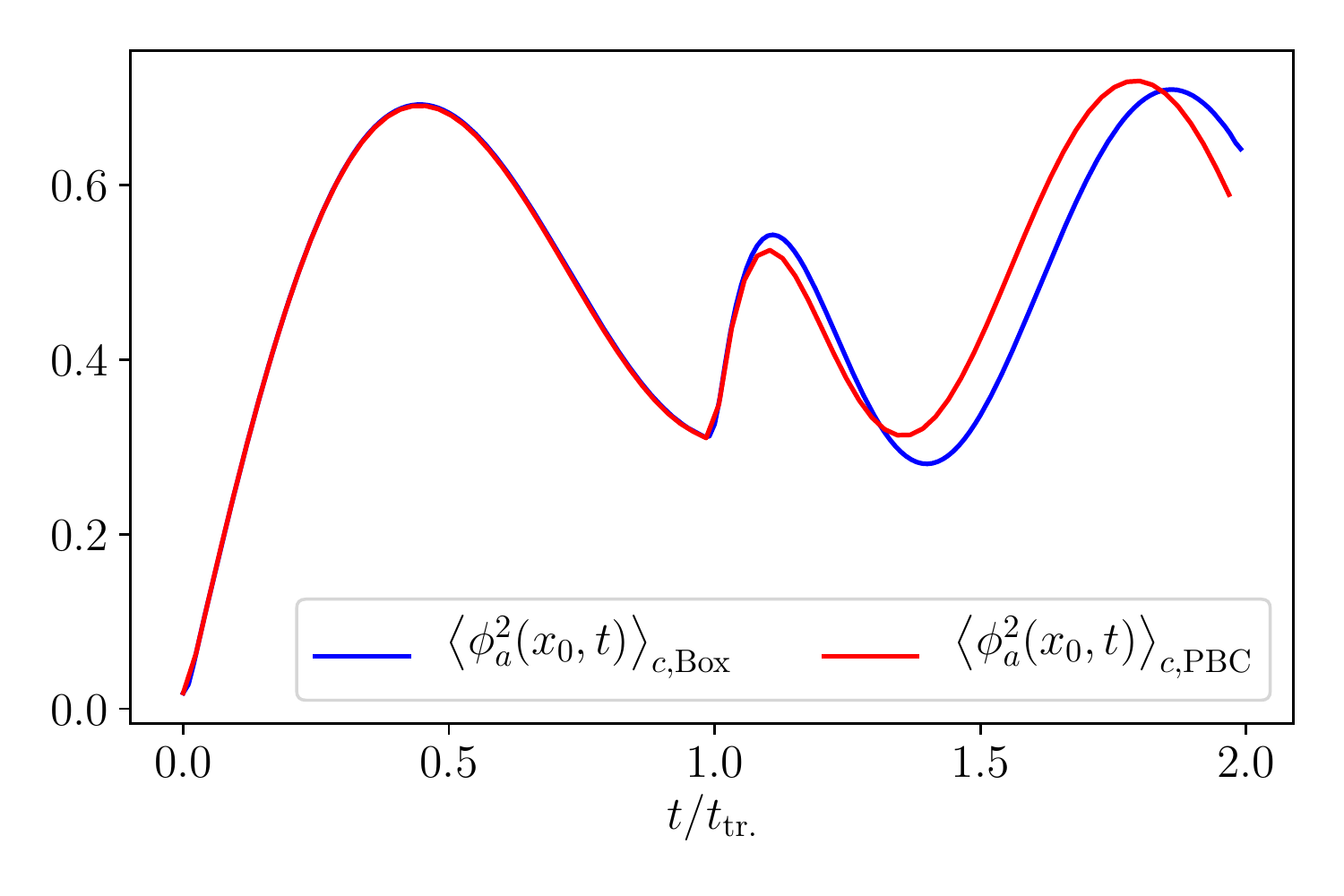}
	\caption{Comparison between results for box boundary conditions (blue) and periodic boundary conditions (red). The curves are in perfect agreement until the traversal time $t_{\mathrm{tr}}=L/2v$, after which deviations occur. (a) Josephson oscillations of phase (radians) at the centre of the gas, $x_{0} = L/2$. (b) initial growth and subsequent oscillations in the variance of the relative phase.}
	\label{fig:BC_compare}
\end{figure}

To connect with our previous work \cite{Nieuwkerk2018b} we include a comparison between results from that paper, where periodic boundary conditions were used, and the results derived for a box geometry in the present paper. Fig. \ref{fig:BC_compare} shows that the two geometries give extremely similar results in the centre of the trap for times below the traversal time, whereas deviations do occur after this time. It should also be noted that in \cite{Nieuwkerk2018b} and Fig. \ref{fig:BC_compare}, results are presented for smaller tunnel couplings ($t_{\perp} \approx 1.17\, \mathrm{Hz}$) than in the rest of this paper. The reason for choosing these values in \cite{Nieuwkerk2018b} was that for a relatively shallow field potential, the inharmonicity of the cosine in the sine-Gordon model manifests itself more strongly, making deviations from the purely quadratic theory more apparent. For the purposes of this paper, however, it is more interesting to look at relatively large tunnel-couplings ($t_{\perp} = 15\, \mathrm{Hz}$, see Sec. \ref{sub:experimental_parameters}), as this enhances the coupling between the sectors in which we are interested.
\subsubsection{\secfix{Finite coupling between sectors ($\sigma>0$)
    and homogeneous initial conditions}}
\begin{figure}[ht]
	\centering
	(a)\includegraphics[width=0.46\textwidth]{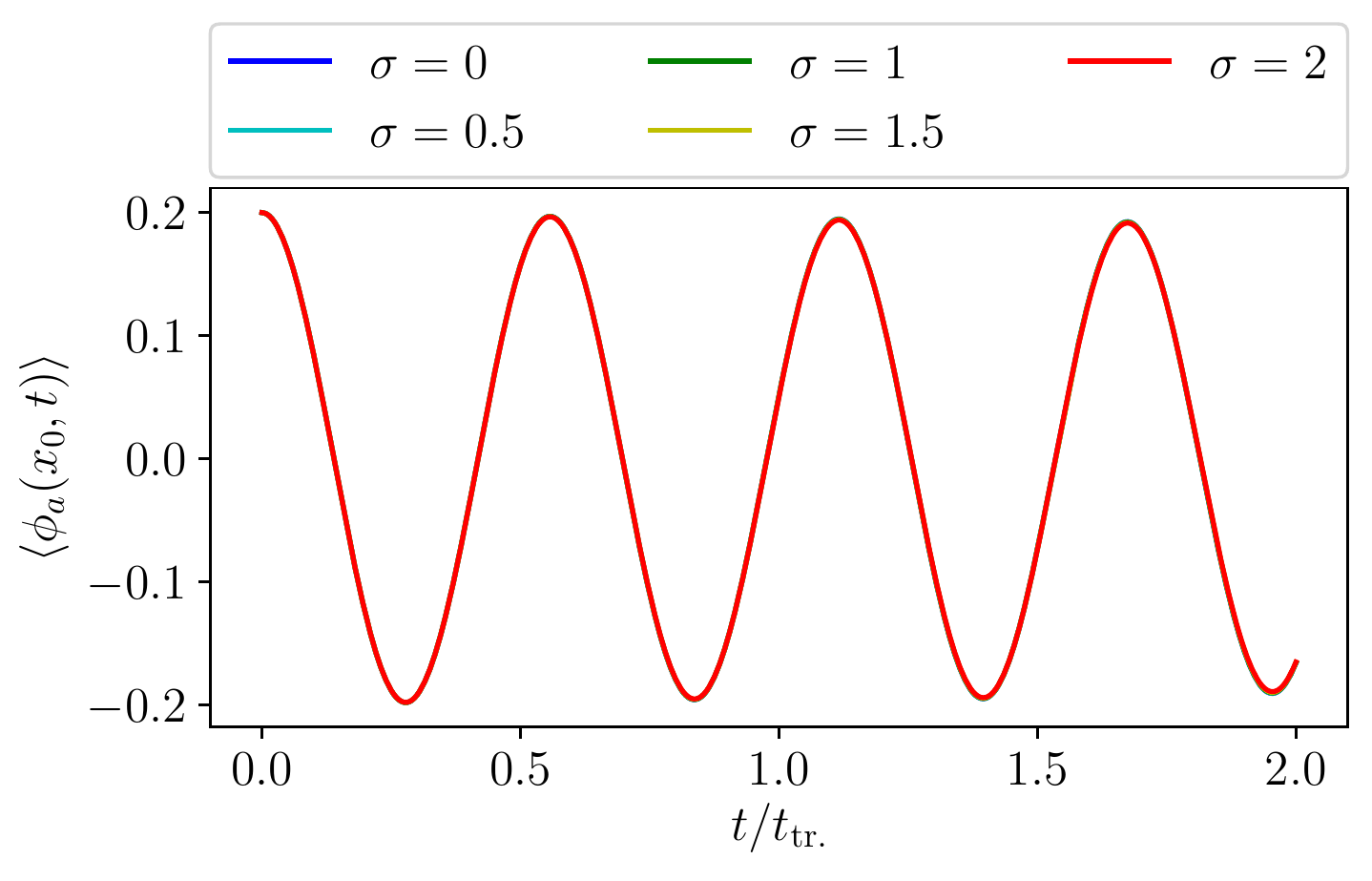}
	(b)\includegraphics[width=0.46\textwidth]{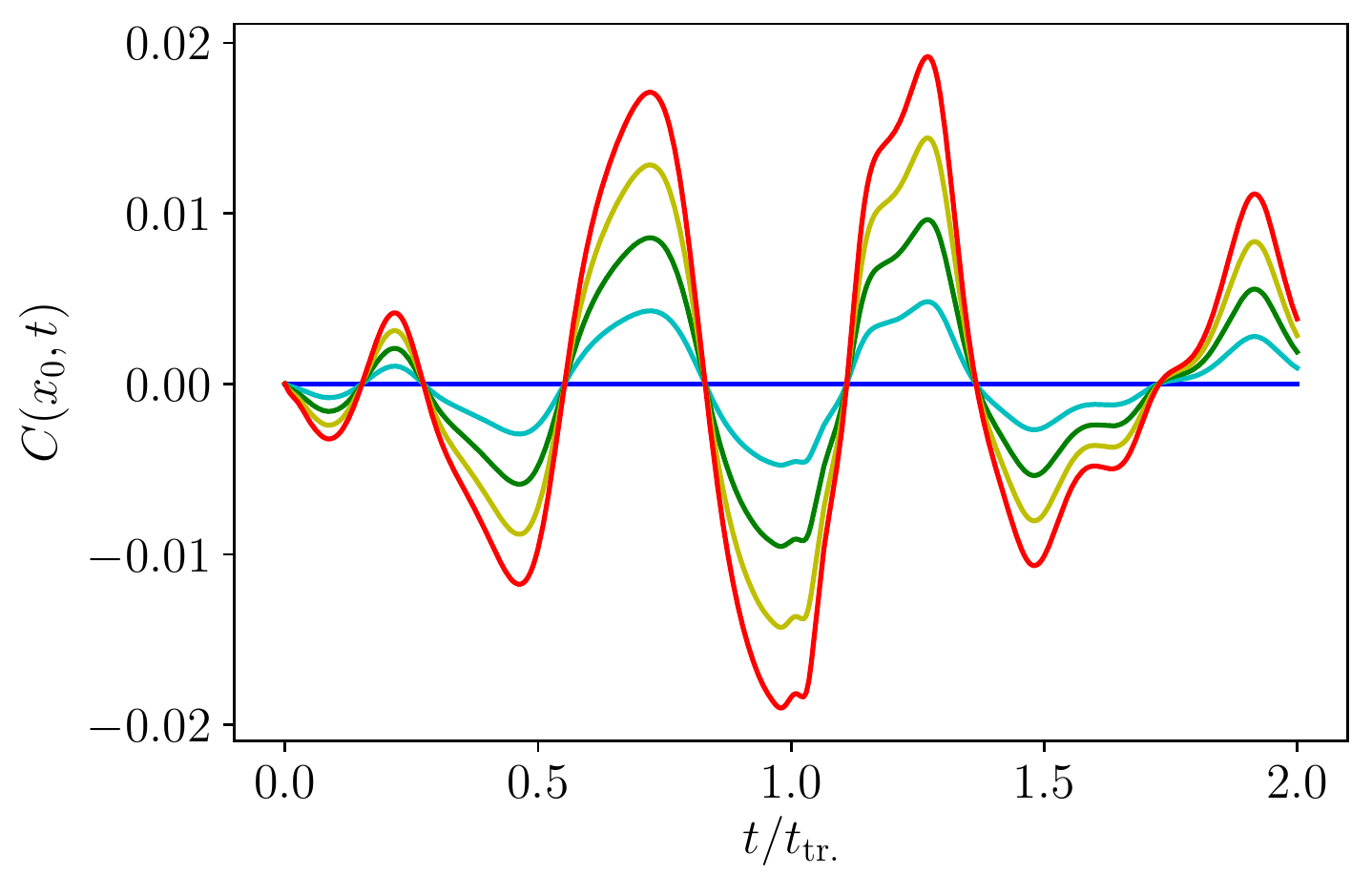}\\
\caption{(a) time evolution of the phase in the antisymmetric sector at the box centre $x_{0}=L/2$. Curves are displayed for
  different values of $\sigma$, with a flat intial density profile
  $\langle\Pi_s(x)\rangle=0$. A change of $\sigma$ has no appreciable effect on this observable.
   (b) a somewhat stronger effect is the development of correlations between
  $\phi_{a,s}$, where the normalized covariance from
  Eq. (\ref{eq:covariance}) is displayed, for $x_{0}=L/2$.}
\label{fig:flat_profile_sigmas}
\end{figure}

\begin{figure}[ht]
	\centering
	\includegraphics[width=0.45\textwidth]{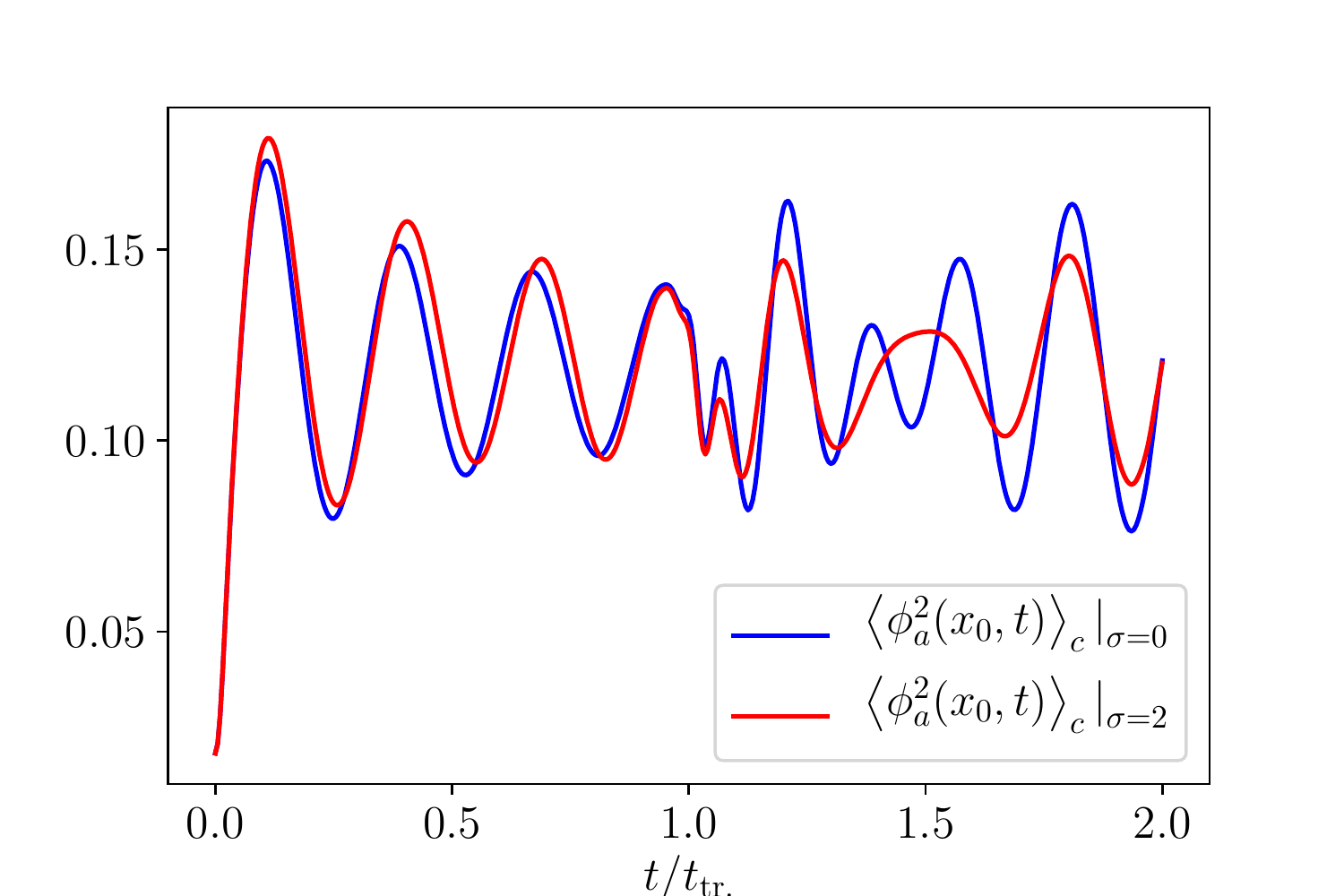}
	\caption{Variance of the relative phase, for $\sigma = 0$ (blue) and $\sigma= 2$ (red). A slight increase in the variance is
          visible for the larger value of $\sigma$ for times below $t_{\mathrm{tr}}=L/2v$} 
	\label{fig:var_phi}
\end{figure}

We next investigate different values of the coupling constant
$\sigma$, and the resulting mixing between the
sectors. Fig. \ref{fig:flat_profile_sigmas} shows results for
$\sigma=0,1/2,1,3/2,2$, starting from completely flat profiles, as in
Eqs. (\ref{eq:profiles_flat_a}), (\ref{eq:profiles_flat}). When
increasing $\sigma$, the phase oscillations remain essentially
unchanged. A stronger effect is visible in the covariance between
$\phi_{a}$ and $\widetilde{\phi_{s}}$, however. To quantify this, we
define 
\begin{align}
C(x,t) \equiv \frac{\big< \widetilde{\phi_{s}}(x,t) \phi_{a}(x,t)
  \big>_{c} }{\sqrt{\big< \widetilde{\phi_{s}}(x,t)
    \widetilde{\phi_{s}}(x,t) \big>_{c} \big< \phi_{a}(x,t)
    \phi_{a}(x,t) \big>_{c}}}\ . \label{eq:covariance} 
\end{align}
As can be seen in Fig. \ref{fig:flat_profile_sigmas}(b), the
covariance $C(x,t)$ increases to appreciable values as $\sigma$ is
increased. We also note that for larger values of $\sigma$, the
variance of the relative phase increases somewhat for times below the traversal time, see
Fig. \ref{fig:var_phi}. 

\begin{figure}[ht]
	\centering
	(a)\includegraphics[height=0.37\textwidth]{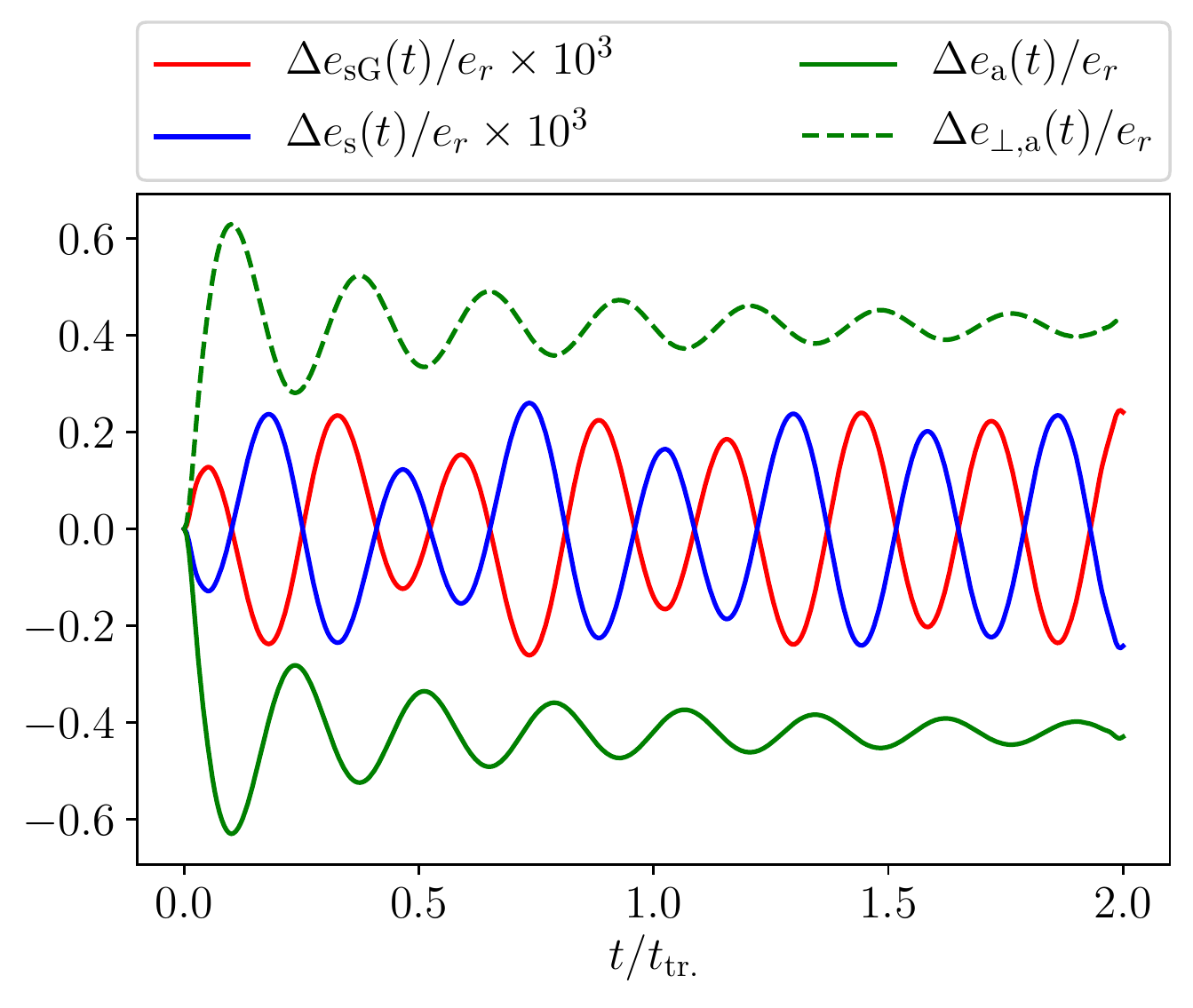}
	(b)\includegraphics[height=0.37\textwidth]{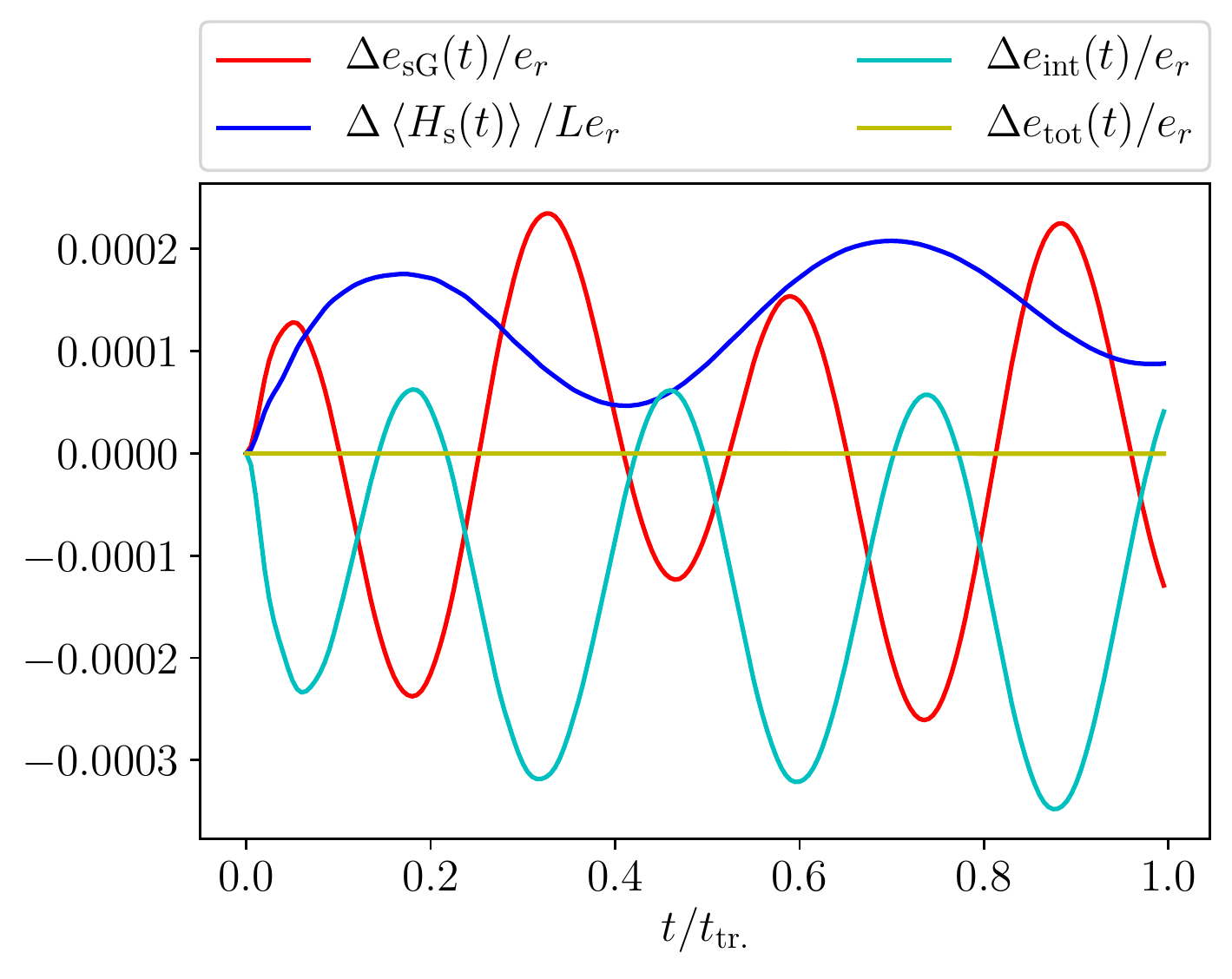}
	\caption{Energy flow between the different terms in Eqs. (\ref{eq:energies4}), as
          a ratio with the reference scale $e_{r} = \braket{H_{s}(0)}/L$.} 
	\label{fig:E_flat}
\end{figure}

It is also instructive to consider the energy flow between different
terms in the Hamiltonian. To this end we define the following quantities
\begin{align}
e_{a,0}(t) &= \frac{\braket{H_{a}}}{L}\ , \quad
e_{a, \perp}(t) = - \frac{2 t_{\perp} \rho_{0}}{L} \int_{0}^{L} dx \,
\braket{\cos \phi_{a}(x)}\ ,\quad
e_{sG}(t) = e_{a,0}(t) + e_{a, \perp}(t)\ , \nn
e_{\rm int}(t) &= - \frac{2 T_{\perp} \sigma}{L}
\int_{0}^{L} dx \, \braket{\Pi_{s}(x) \cos
  \phi_{a}(x)}, \quad e_{s}(t) = e_{\mathrm{int}}(t) + \braket{H_{s}(t)}/L\ .\label{eq:energies4} 
\end{align}
We note that the total energy density, which is given by $e_{sG}(t) +
e_{\rm int}(t)+\langle H_s\rangle/L$, is independent of time, as
required for a closed quantum system. Since we are interested in the time
dependence of the various energy densities we subtract their values in
the initial state and consider
\begin{align}
\Delta e_{j}(t) \equiv e_{j}(t) - e_{j}(0)\ .
\end{align}
To quantify the effects of the $\sigma$-coupling on the flow of
energy from and to the sine-Gordon model we show $\Delta e_{\rm
  SG}(t)$ in Fig. \ref{fig:E_flat}. To ascertain which fraction
of the energy change is due to the kinetic and interaction parts of
the sine-Gordon model we also show $\Delta e_a(t)$ and $\Delta
e_{\perp,a}(t)$ in Fig. \ref{fig:E_flat}(a). We observe that the change in $\Delta e_{\rm
  SG}(t)$ is very small, as significantly larger changes in $\Delta e_a(t)$ and $\Delta
e_{\perp,a}(t)$ largely compensate each other. In Fig. \ref{fig:E_flat}(b) we show how much of the energy from the sine-Gordon model $\Delta e_{\rm
  SG}(t)$ ends up in the new interaction term $e_{\mathrm{int}}(t)$ and how much goes to $\braket{H_{s}(t)}/L$.
      
\subsubsection{\secfix{Finite coupling between sectors ($\sigma>0$) and
inhomogeneous initial conditions}}
\begin{figure}[ht]
\centering
(a)\includegraphics[width=0.45\textwidth]{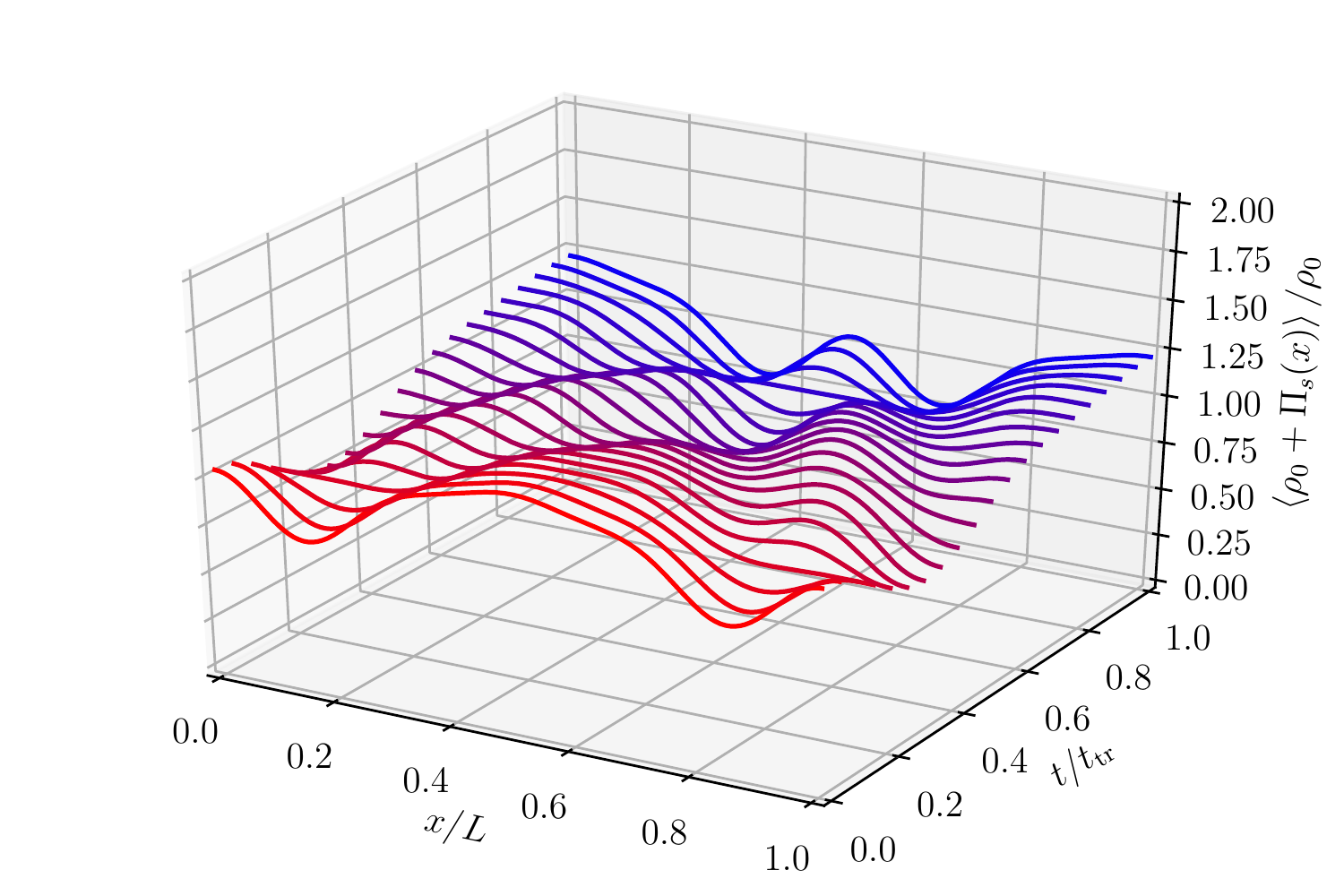}
(b)\includegraphics[width=0.45\textwidth]{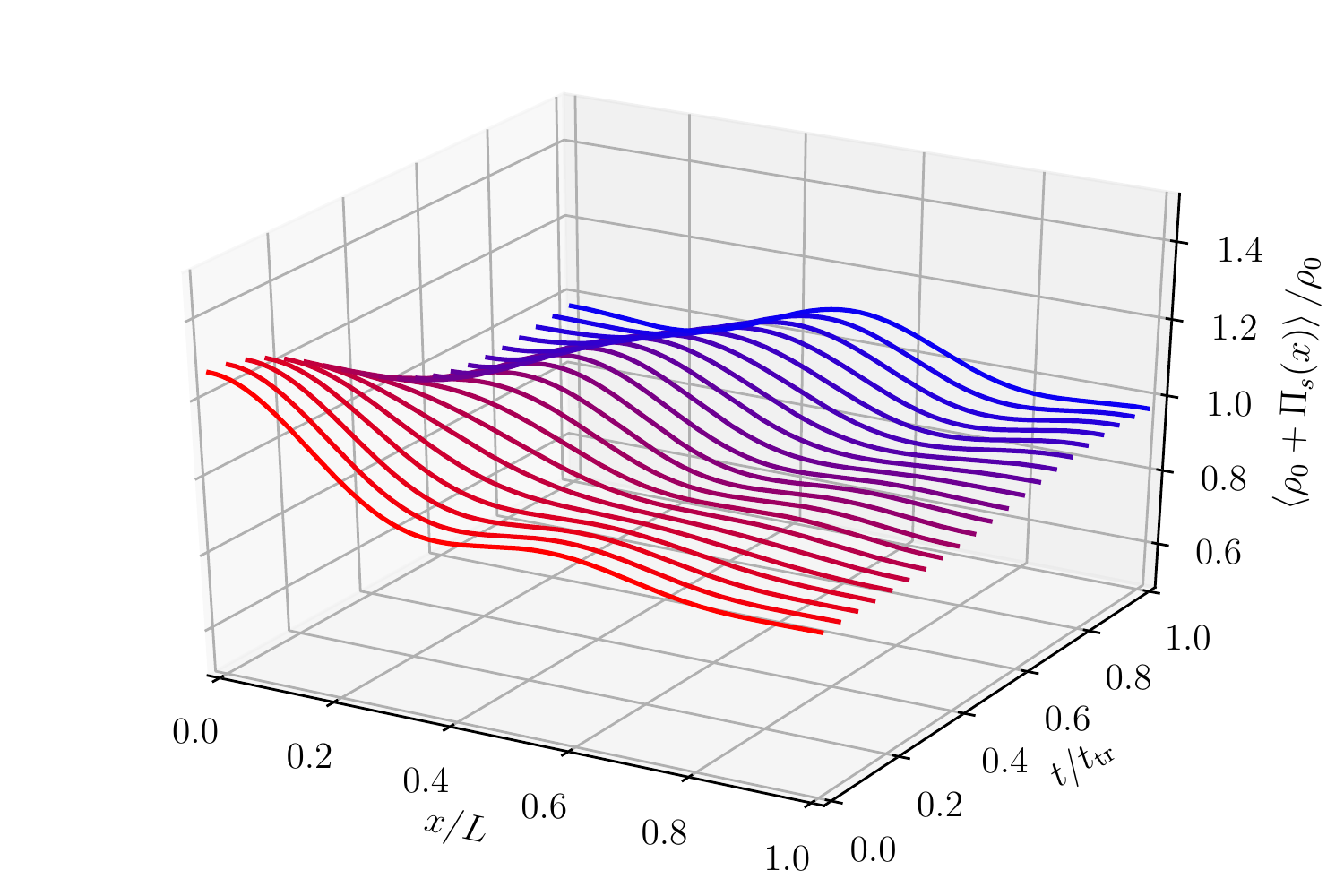}
\caption{Examples of the time evolution of the density profile for $\sigma=0$. The initial profile in (a) is
  symmetric around the origin, while the one in (b) is not.}
\label{fig:Pi_profiles}
\end{figure}

\begin{figure}[ht]
\centering
(a)\includegraphics[width=0.45\textwidth]{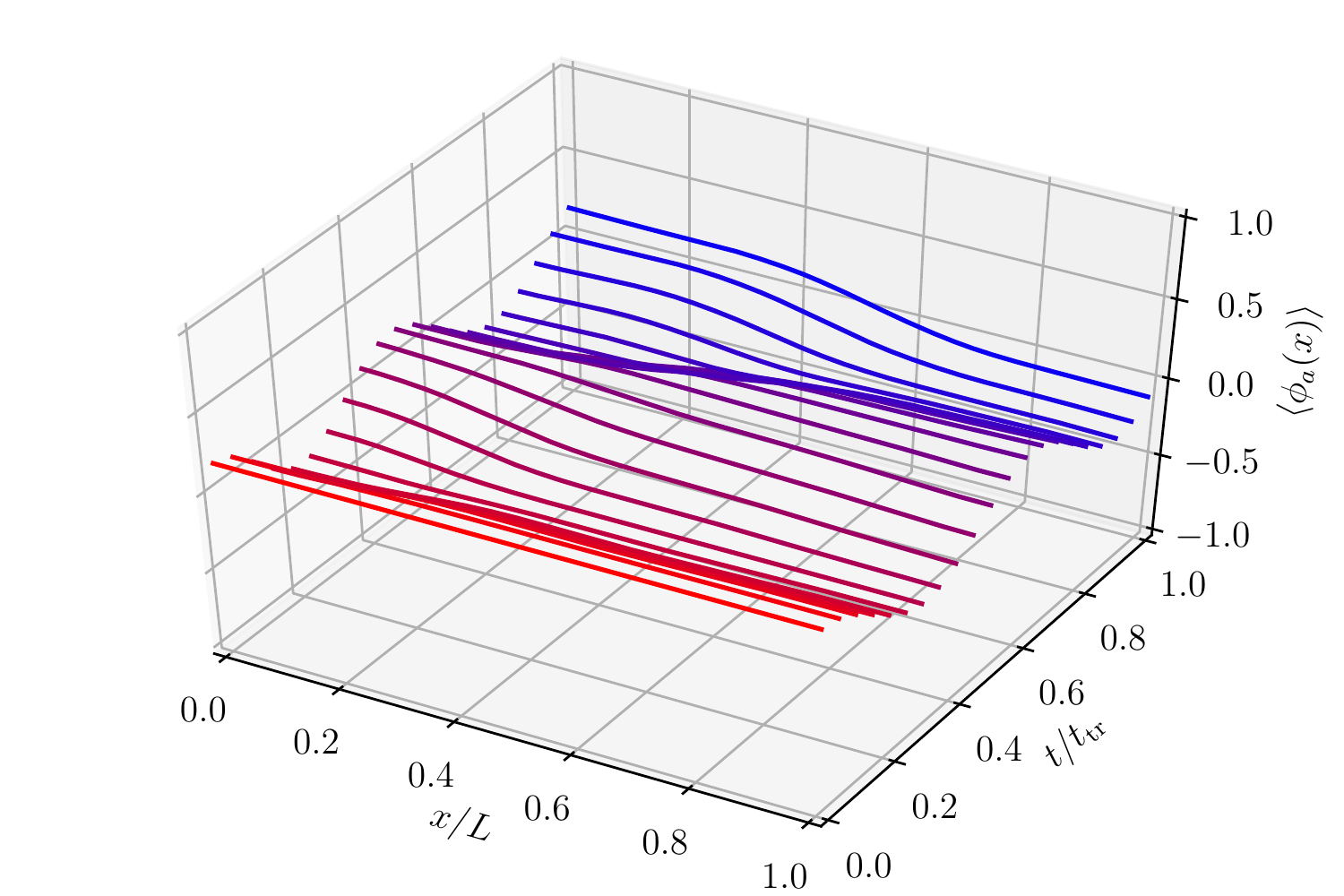}
(b)\includegraphics[width=0.45\textwidth]{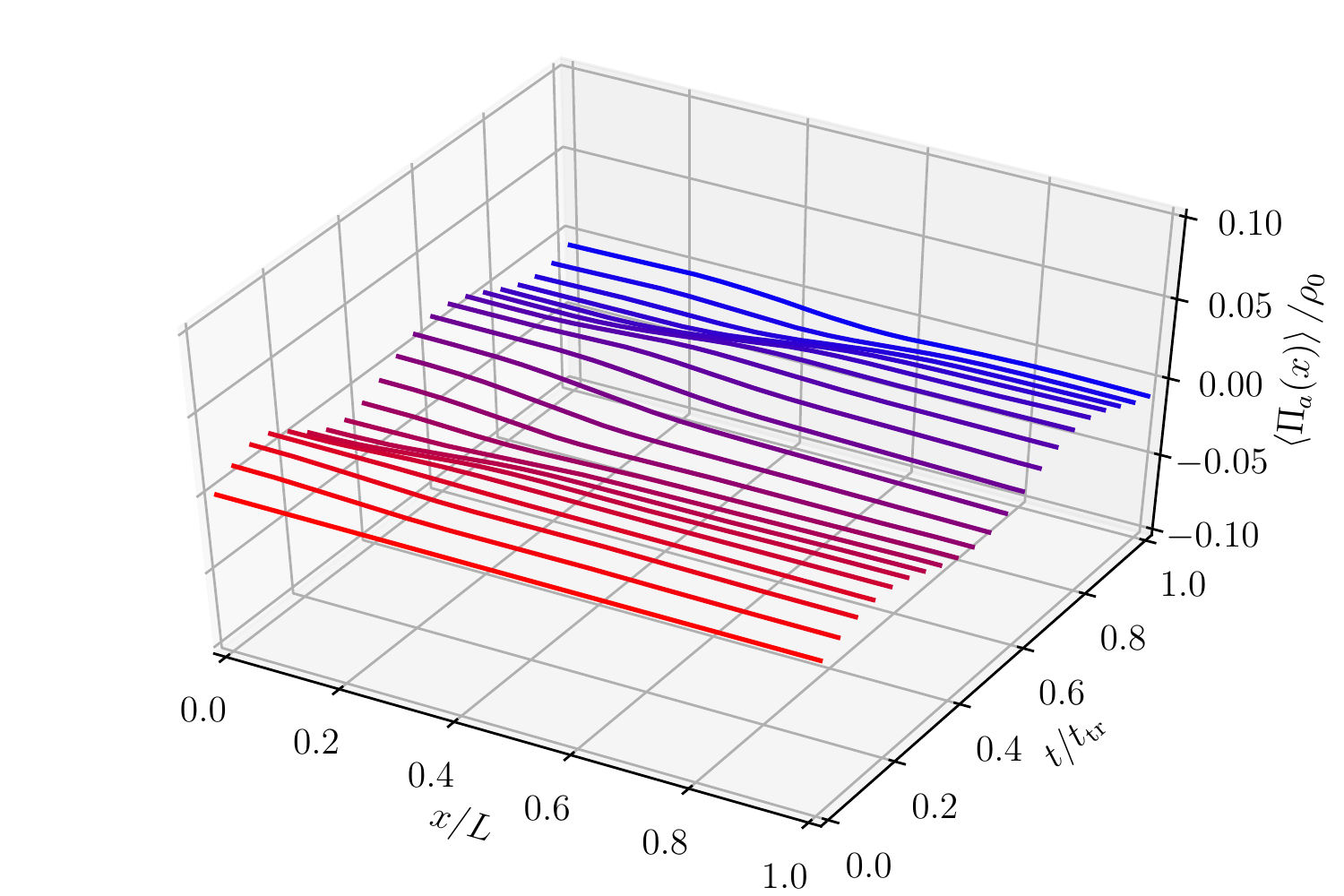}
\caption{(a) The time and position dependence of $\braket{\phi_{a}(x)}$ corresponding to
the same initial condition as Fig.~\ref{fig:Pi_profiles}(a) with $\sigma=2$. We see
that the initially flat profile develops inhomogeneities due to the
sector coupling. (b) the same as panel (a), but showing $\braket{\Pi_{a}(x)}$.}
\label{fig:profiles_non_flat}
\end{figure}

\begin{figure}[ht]
	\centering
\includegraphics[width=0.48\textwidth]{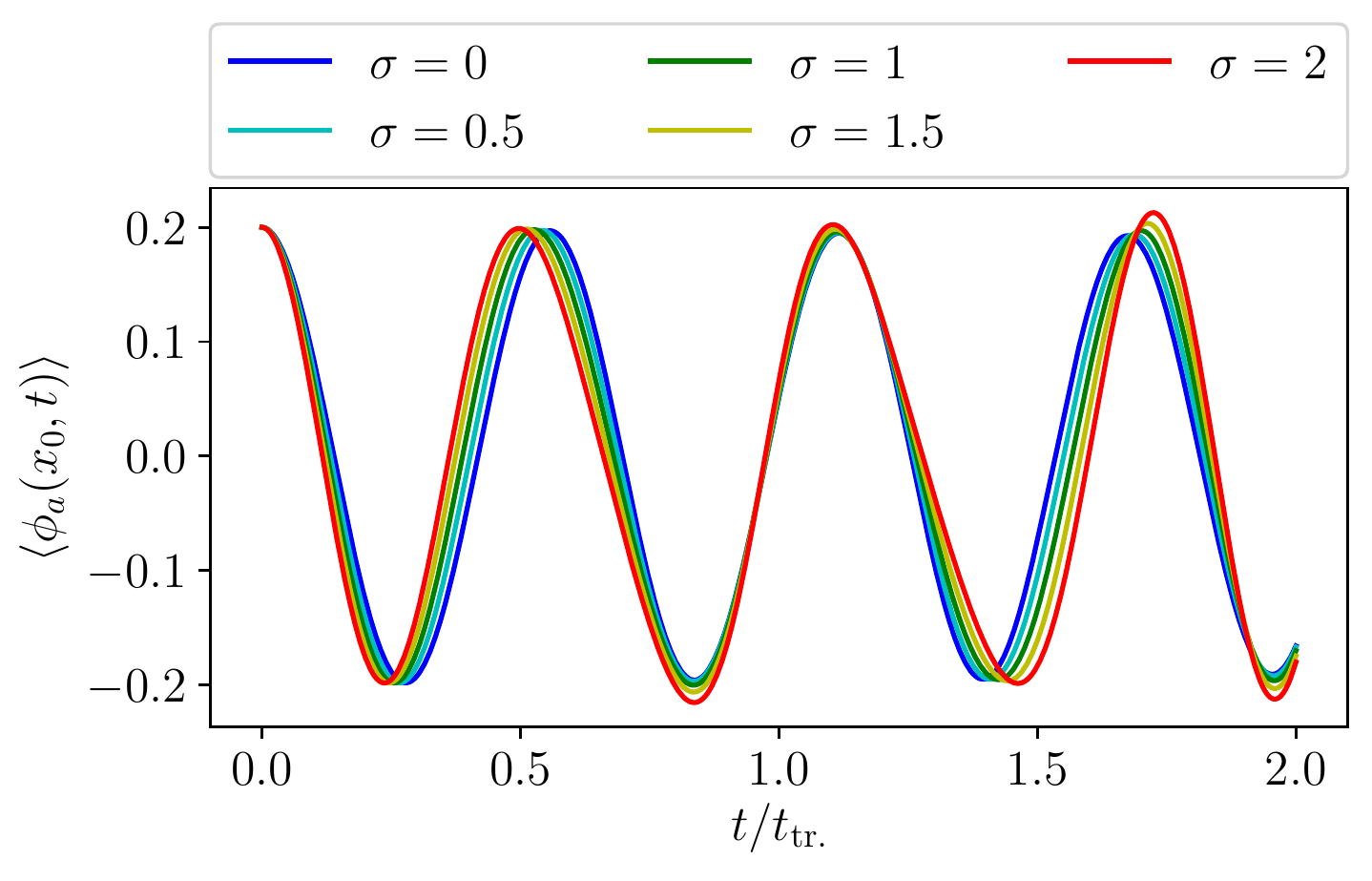}
\quad
\includegraphics[width=0.48\textwidth]{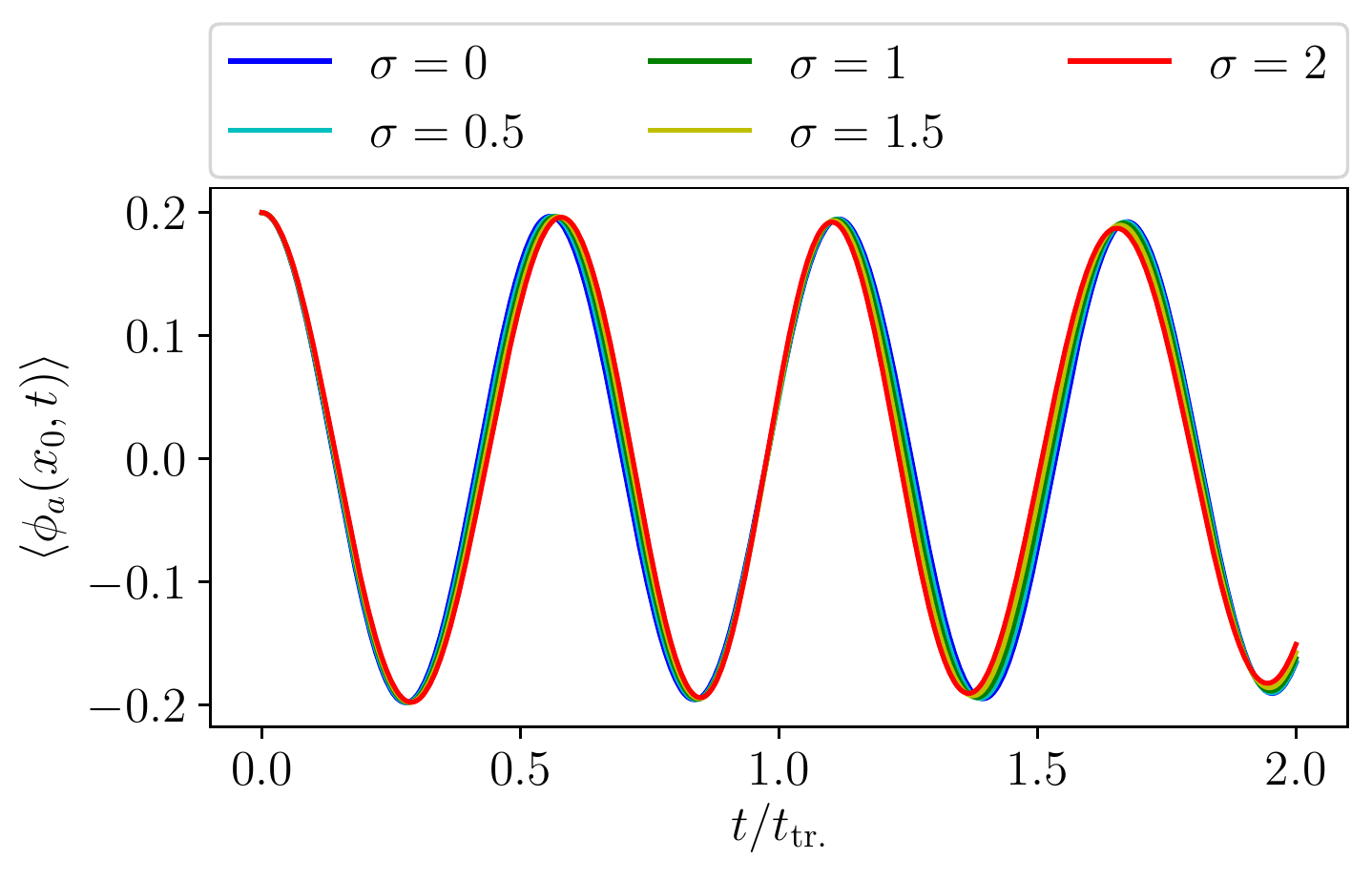}
\caption{Time dependence of the relative phase in the centre of the box for
the same initial conditions as Fig.~\ref{fig:Pi_profiles}(a) and (b), respectively.}
	\label{fig:phi_midpoint_nonflat}
\end{figure}

\begin{figure}[ht]
	\centering
\includegraphics[width=0.48\textwidth]{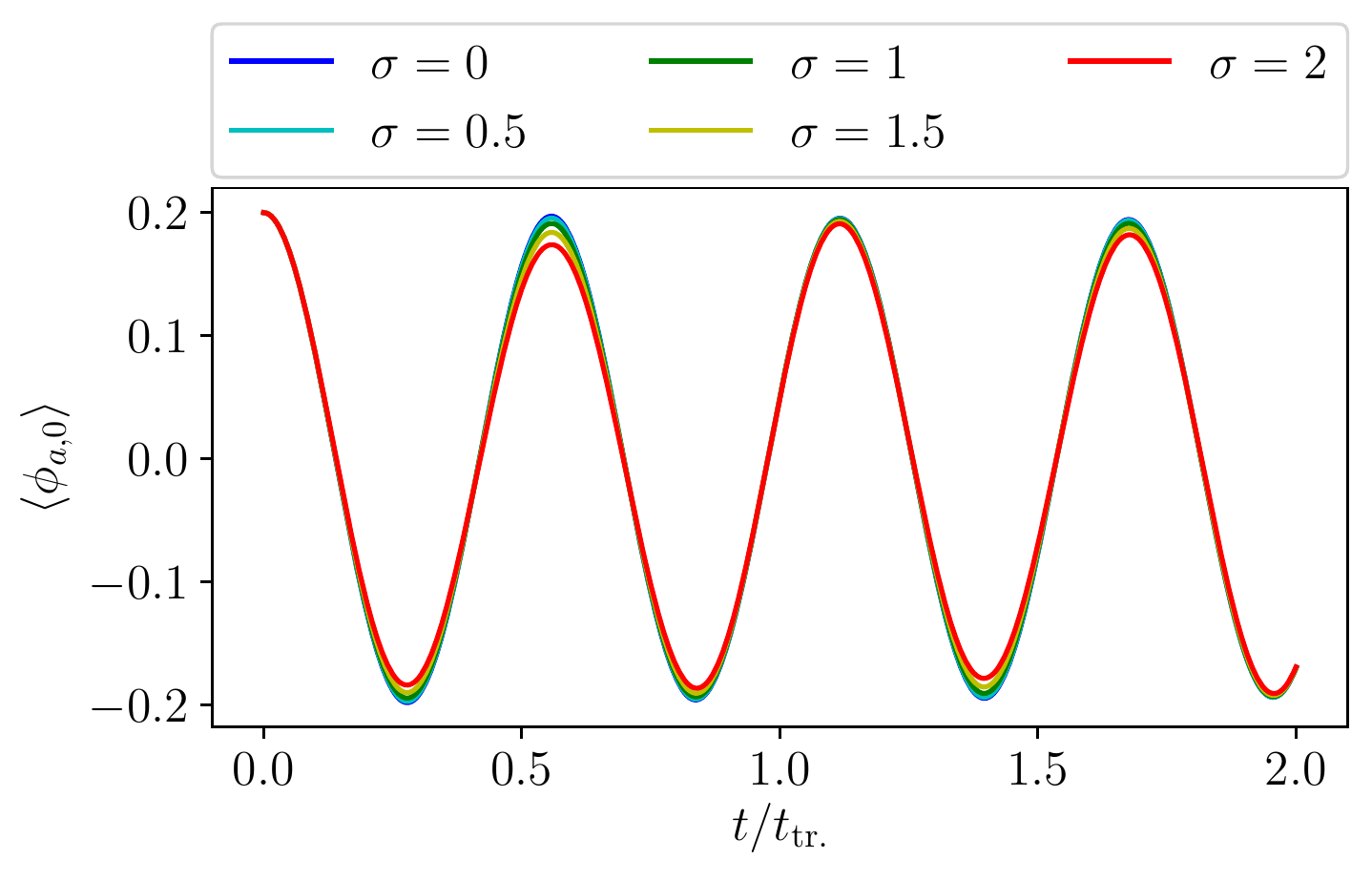}
\quad
\includegraphics[width=0.48\textwidth]{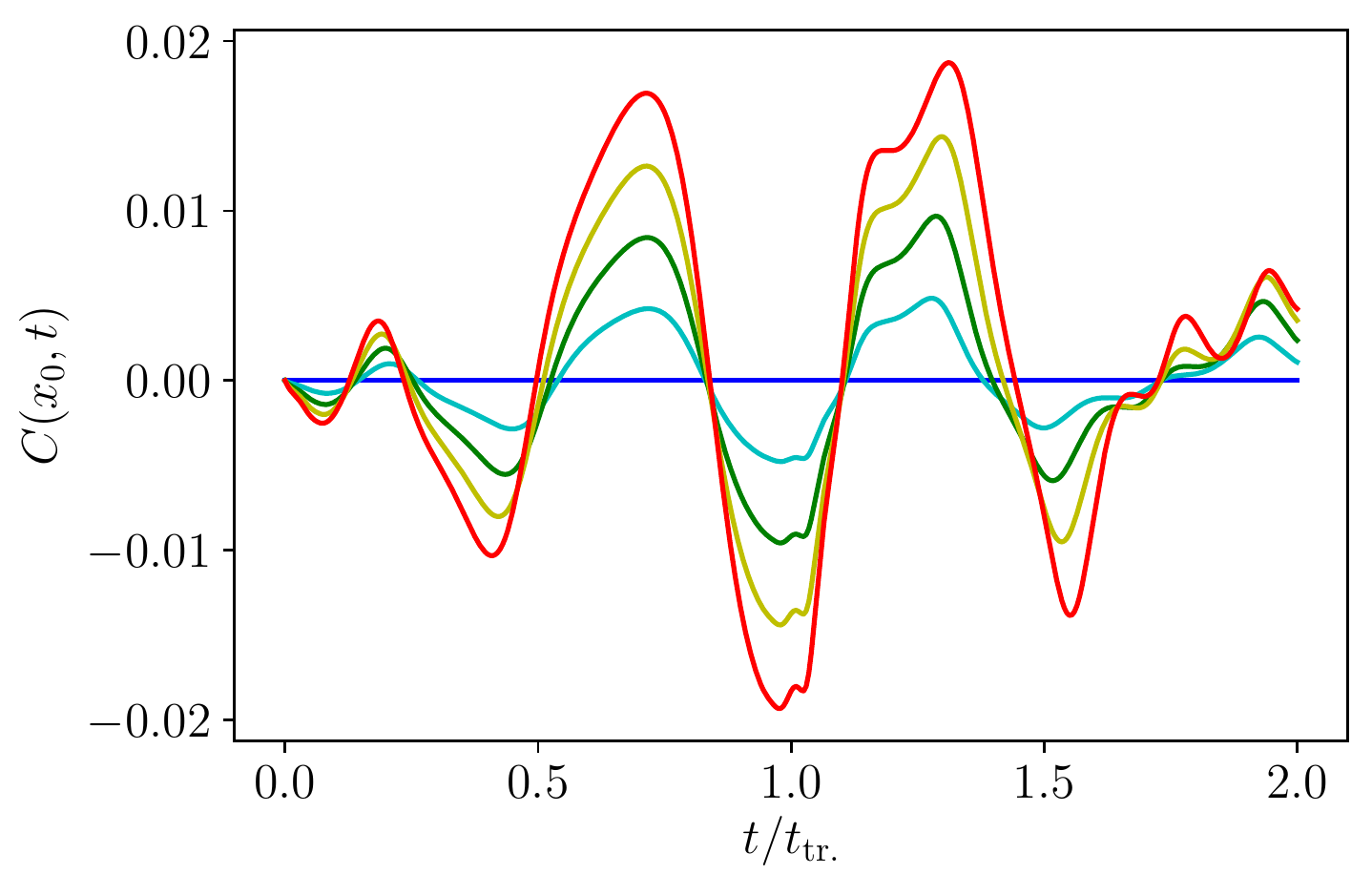}
\caption{Time dependence of the relative phase and the covariance
  $C(x_0,t)$ \fr{eq:covariance} in the centre of the box for 
the profiles shown in panel (a) of Fig. \ref{fig:Pi_profiles}.}
	\label{fig:ZM_cov_nonflat1}
\end{figure}

\begin{figure}[ht!]
	\centering
\includegraphics[width=0.48\textwidth]{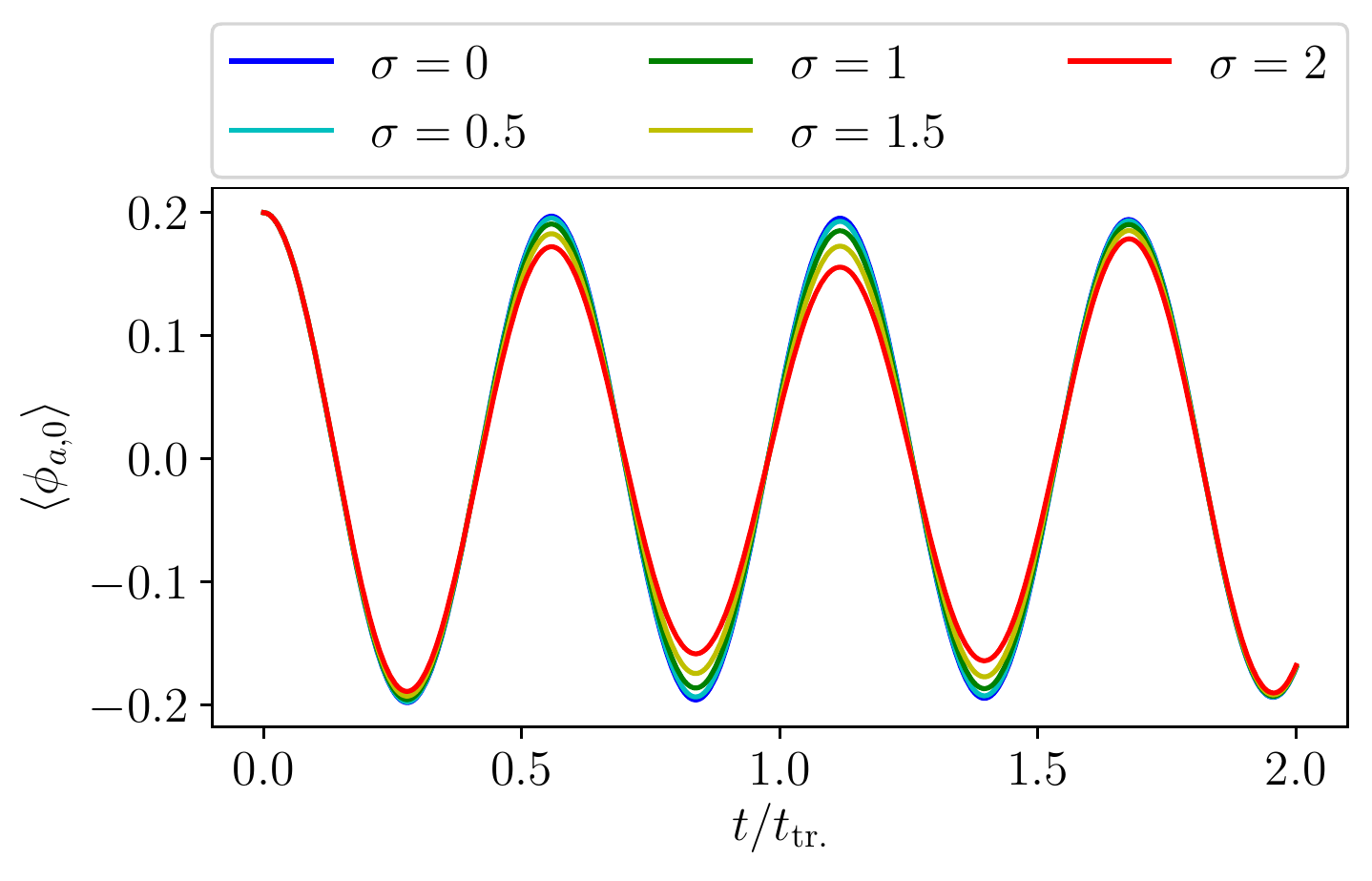}
\quad
\includegraphics[width=0.48\textwidth]{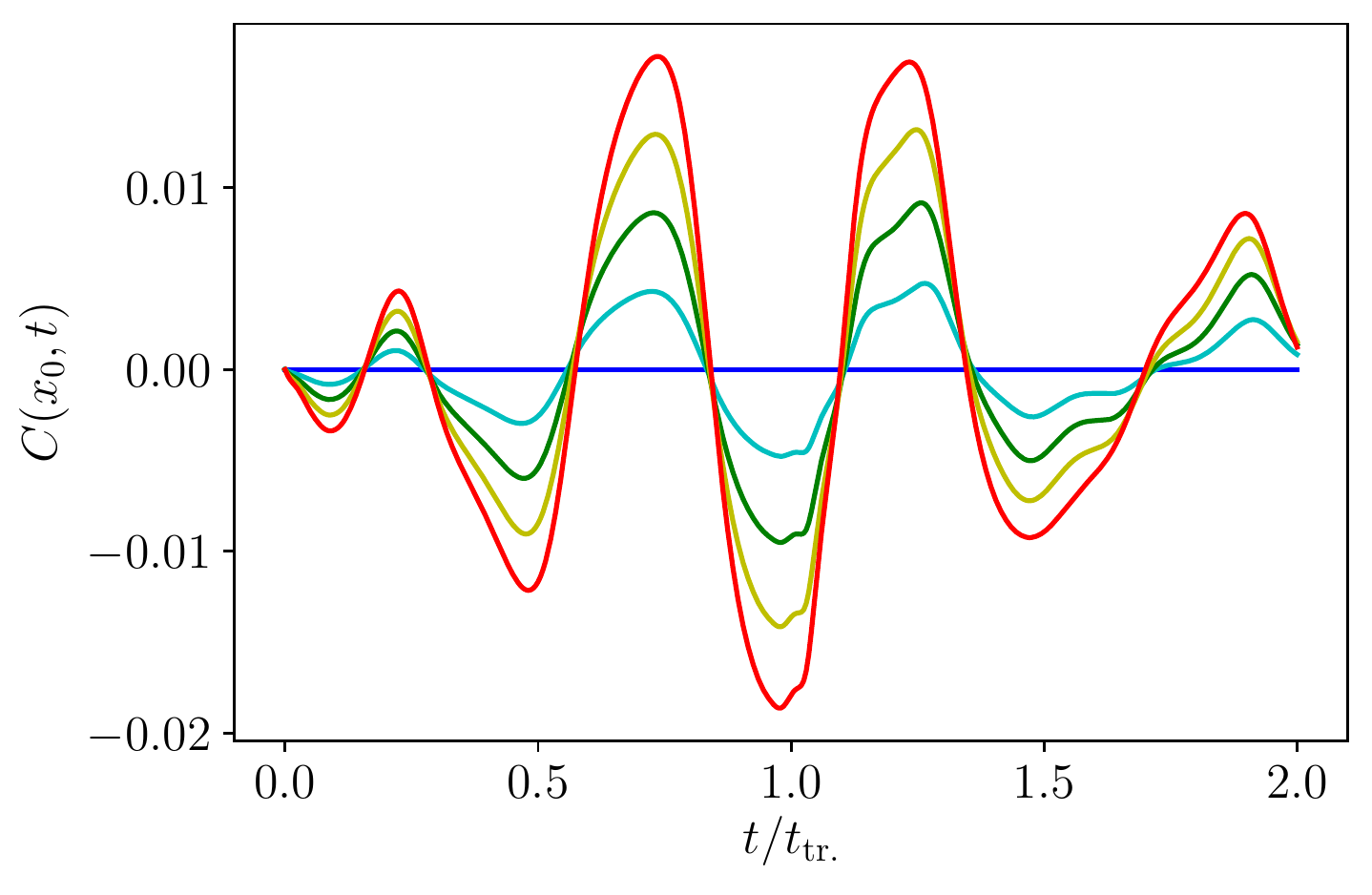}
\caption{Time dependence of the relative phase and the covariance
  $C(x_0,t)$ \fr{eq:covariance} in the centre of the box for 
  the profiles shown in panel (b) of Fig. \ref{fig:Pi_profiles}.}
  \label{fig:ZM_cov_nonflat2}
\end{figure}

As a next step, we investigate the effect of initial density
profiles $\braket{\Pi_{s}(x)}$ that are spatially inhomogeneous. These
profiles will evolve in time as is shown in
Fig. \ref{fig:Pi_profiles} (a,b).
The profiles $\braket{\phi_{a}(x)}$ and $\braket{\Pi_{a}(x)}$ are
strongly affected by the strength of the $\sigma$-coupling to the
inhomogeneous profile $\braket{\Pi_{s}(x)}$ and develop
inhomogeneities as a consequence. This is illustrated in
Figs. \ref{fig:profiles_non_flat}(a,b) and has repercussions
for the Josephson oscillations.
The latter now displays spatial variations, which are caused by an effective Josephson frequency that has become $\sigma$- and position-dependent due to the presence of the space-dependent $\Pi_{s}(x)$-field in the interaction term. This local and $\sigma$-dependent Josephson frequency is illustrated in Fig. \ref{fig:phi_midpoint_nonflat}. 
The spatial average of the phase, which is equal to the zero mode
$\phi_{a0}$, does not show any $\sigma$-dependence in its Josephson frequency, see
Figs. \ref{fig:ZM_cov_nonflat1},\ref{fig:ZM_cov_nonflat2}. In this case, however, a $\sigma$-dependent modulation in the amplitudes is visible: the Josephson oscillations at
different points in the box move out of phase due to the spatially
varying Josephson frequency mentioned above. This leads to a decrease
in the spatial average. 

\begin{figure}[ht!]
	\centering
	\includegraphics[height=0.38\textwidth]{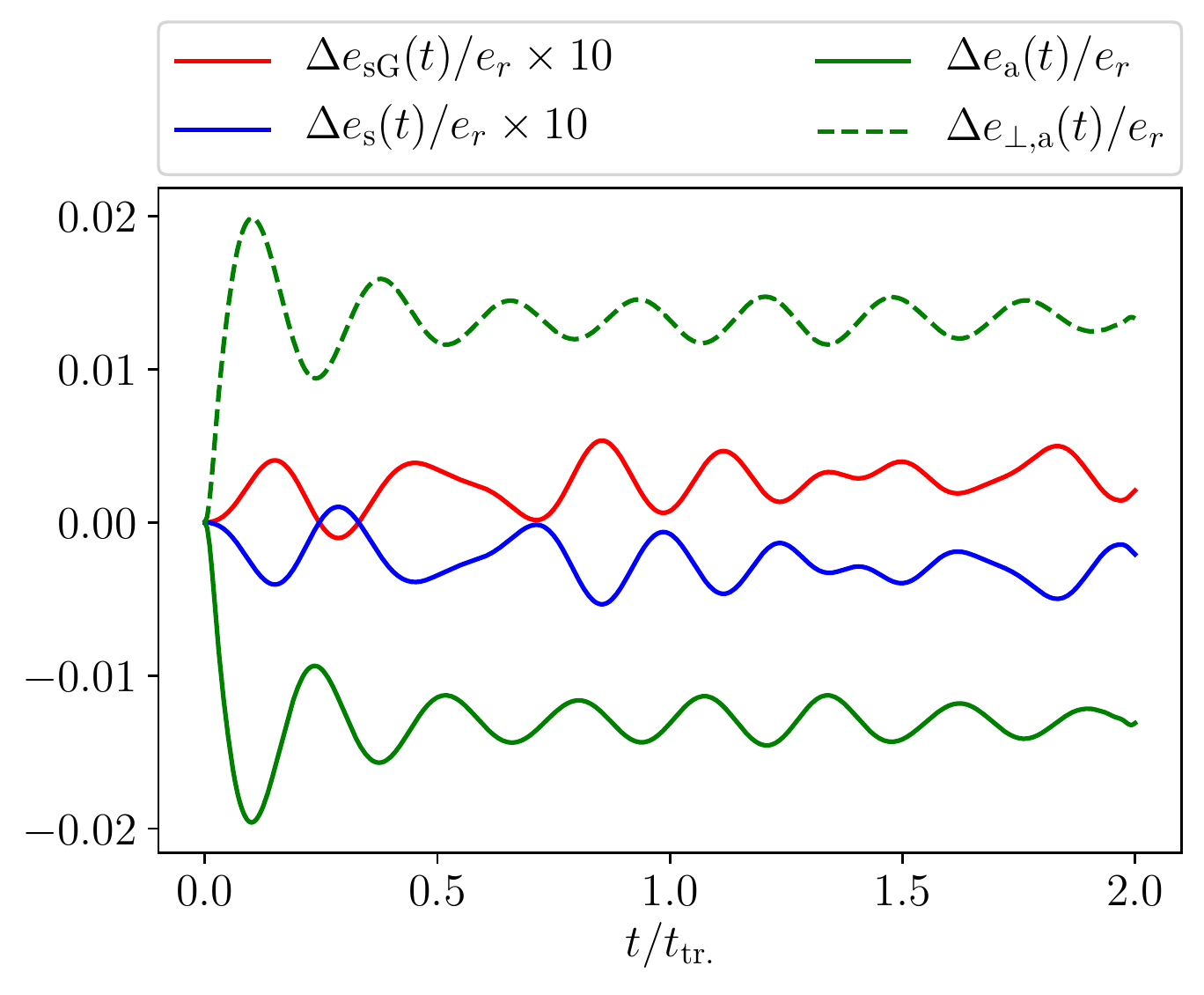}
	\includegraphics[height=0.38\textwidth]{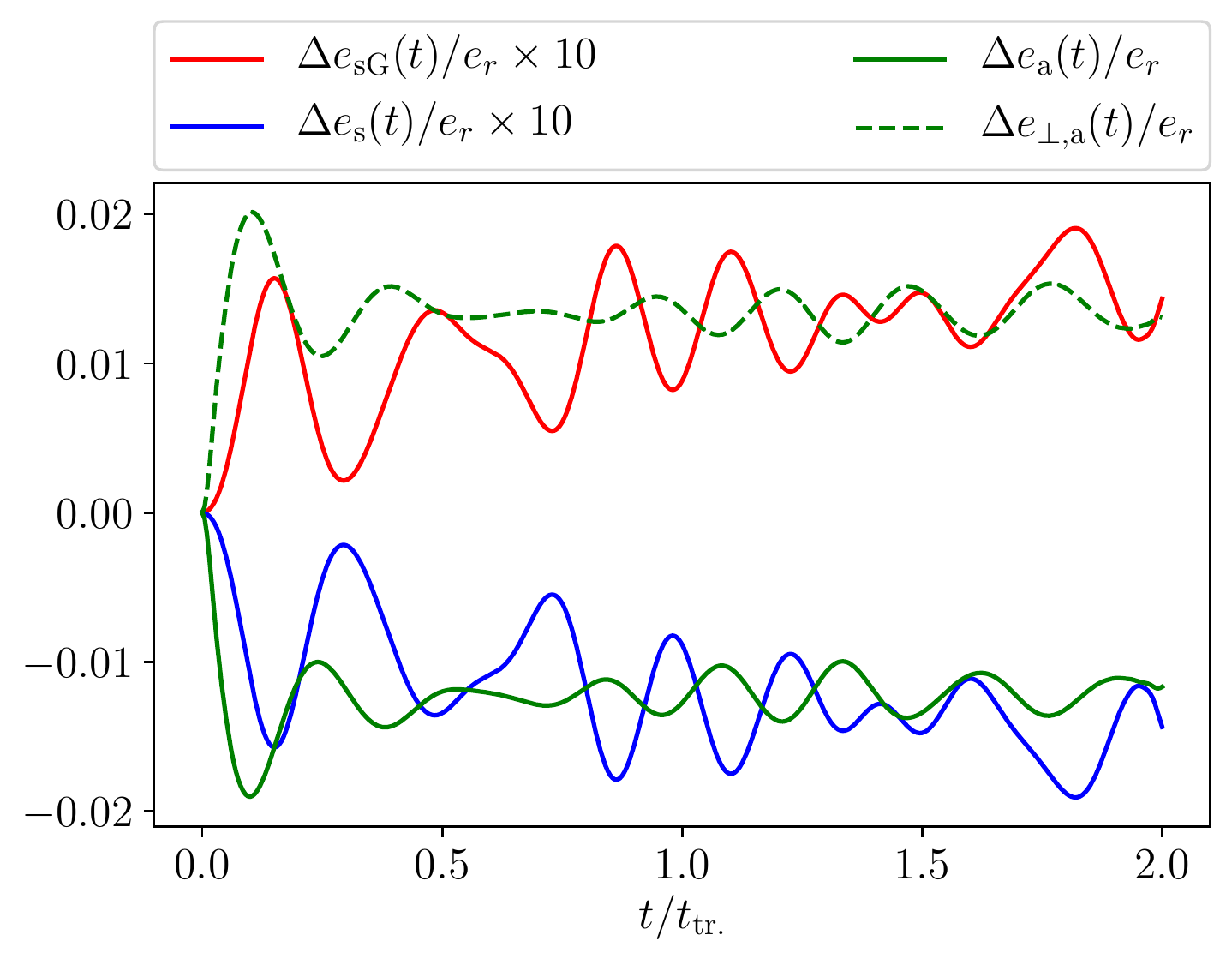}
	\caption{Energy flow between different terms in
          Eqs. (\ref{eq:energies4}), as a ratio with the reference
          scale $E_{r} = \braket{H_{s}(0)}$. Results are shown for the
          density profile from Fig. \ref{fig:Pi_profiles}(a), with
          $\sigma = 1$ (left panel) and $\sigma = 2$ (right panel).}
	\label{fig:var_energy_flows_compare}
\end{figure}

For an inhomogeneous profile of $\braket{\Pi_{s}(x,0)}$, the
covariance grows in time, in resemblance with the homogeneous
case. This happens to an extent that is roughly proportional to
$\sigma$. The same can be said of the energy flow between the
(anti)symmetric sectors, as shown in
Fig. \ref{fig:var_energy_flows_compare}. We see that the
effects of the sector coupling term become stronger when we increase
$\sigma$, but in the window of applicability of our bosonization based
approach the effects remain small.

\subsubsection{Distribution functions of the density after time of flight}

\begin{figure}[htbp!]
	\centering
	(a)\includegraphics[width=0.36\textwidth]{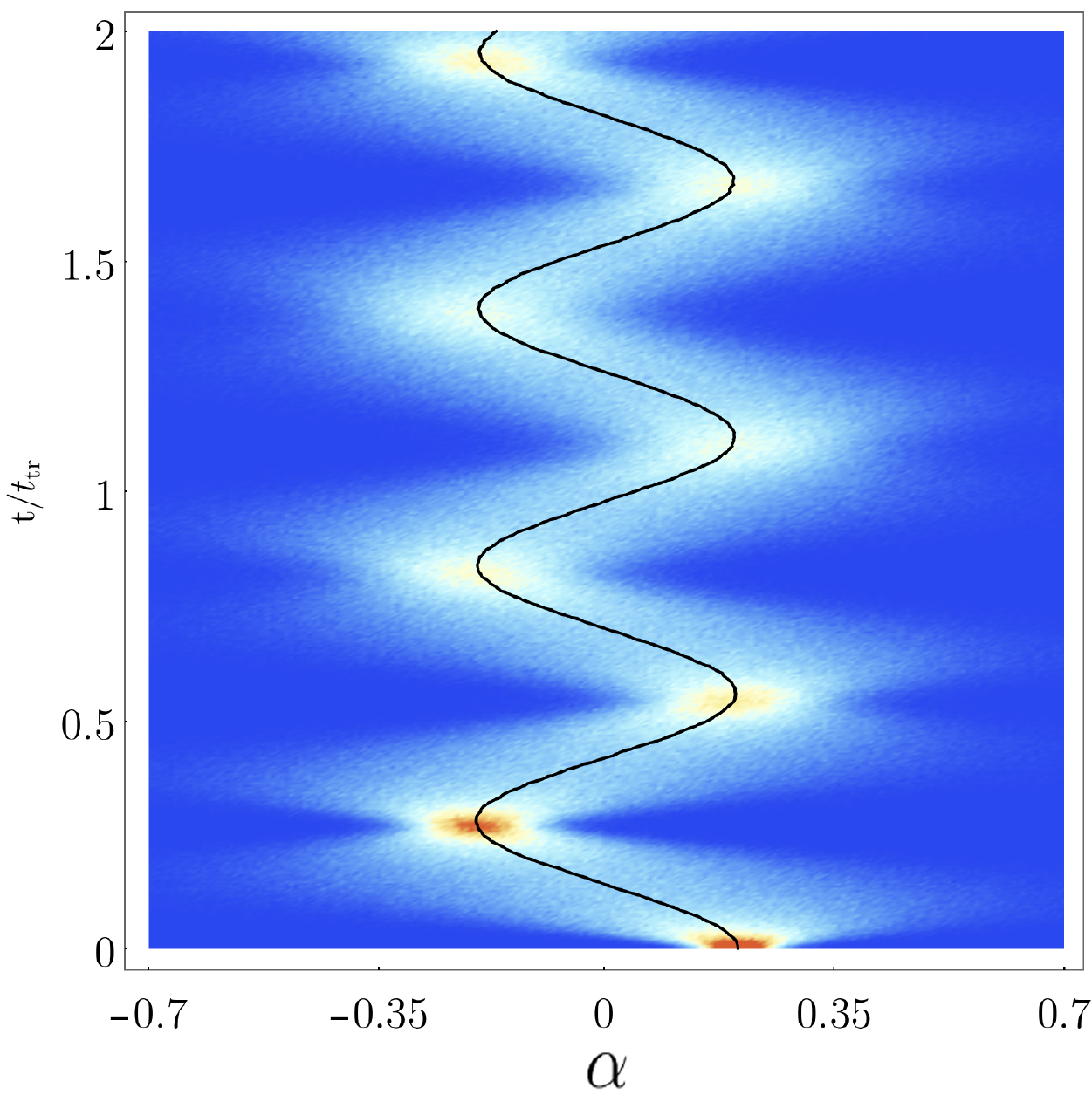}
	\includegraphics[width=0.096\textwidth]{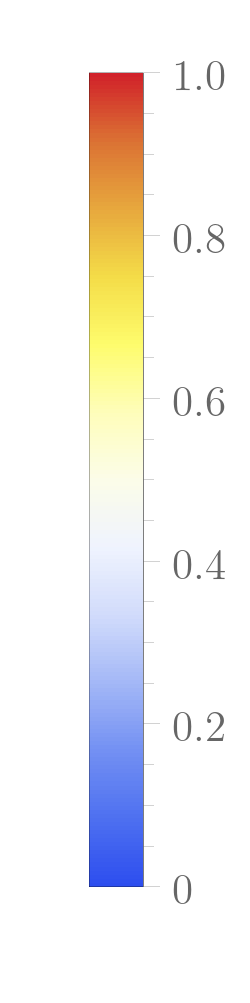}
	(b)\includegraphics[width=0.36\textwidth]{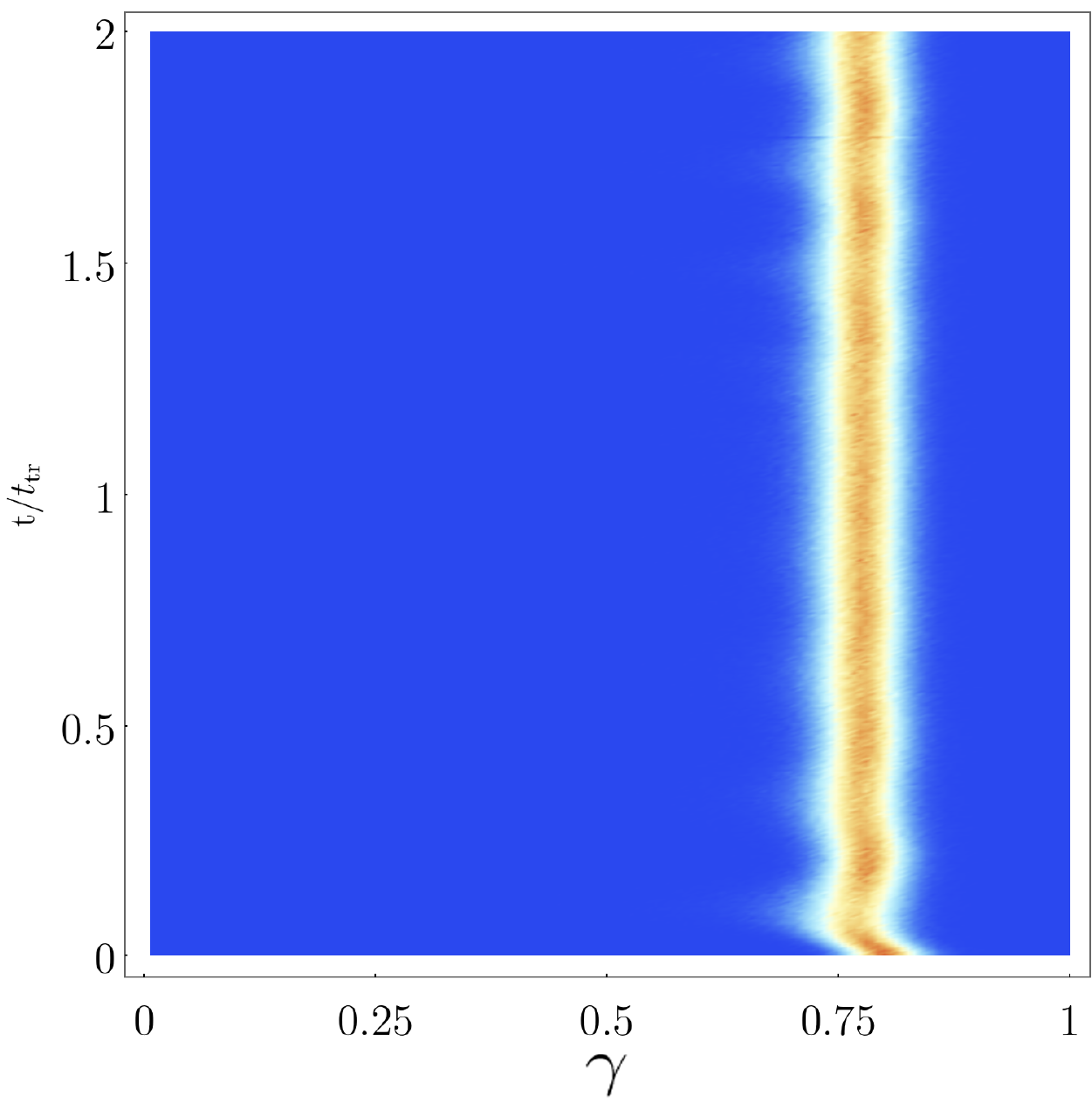}
	\caption{Distribution functions $P_{\Phi_{\ell}}(\alpha,t,t_{1})$ (a) and $P_{C_{\ell}}(\gamma,t,t_{1})$ for the observables $\Phi_{\ell}$ and $C_{\ell}$ defined in Eq. (\ref{eq:interference_eval_parametr}). We choose a time of flight $t_{1} = 15\, \mathrm{ms}$ and integration length $\ell = 20 \, \mu \mathrm{m}$. The density profile used for these plots is homogeneous, with $\sigma = 1$.}
	\label{fig:FDFs}
\end{figure}

As described in Sec. \ref{sub:full_distribution_functions}, our formalism allows the construction of distribution functions for the measured density after time-of-flight expansion. As a proof of principle we present such distribution functions in Fig. \ref{fig:FDFs}, for the observables $\Phi_{\ell}$ and $C_{\ell}$ defined in Eq. (\ref{eq:interference_eval_parametr}).

\section{Conclusion} 
\label{sec:conclusion}

We have extended the theory for non-equilibrium dynamics in pairs of elongated, tunnel-coupled Bose gases using a self-consistent time-dependent harmonic approximation (SCTDHA). In contrast to earlier works, we have studied the effect of a relevant perturbation which couples the (anti)symmetric sectors describing (anti)symmetric combinations of the two Bose gas phases. On top of this, we have dropped the assumtion of translational invariance by placing the system in a box and by imposing inhomogeneous initial density profiles.

Starting from an initial state in which these sectors are
uncorrelated, the coupling of the sectors under time evolution leads
to a number of new but weak effects. First of all we observe the
development of correlations between the sectors over time. This effect
is present for all initial states we have considered, but the
covariance between the sectors never reaches more than a few percent
of the geometric mean of the variances. Second, the spreading of
correlations is accompanied by a small transfer of energy between the
sectors. And finally, the presence of the coupling term makes the
dynamics in the antisymmetric sector susceptible to the breaking of
translational invariance in the symmetric sector. The well-known
Josephson oscillations of relative density and phase are modulated
when taking an inhomogeneous initial density profile of the symmetric
sector. This shows that the role of the trapping potential, which
creates strong inhomogeneities, may play a more important role in
experiment than was previously assumed. However, the model presented
here does not capture the puzzling damping phenomenon observed
recently
\cite{Pigneur2017,schweigler2019correlations,Pigneur2019thesis}. This
is not surprising given that our box potential is very different from
the quadratic potential used in experiment. In future experiments,
however, the box potential is likely to be used, which adds to the
relevance of the calculations presented here. 

We conclude that \textit{(i)} the new term coupling the (anti)symmetric sectors leads to very weak effects. This means that the simulation of a sine-Gordon model using the setup described in this paper should not be severely hampered by the presence of this term. \textit{(ii)} we have shown that it is possible to treat states with broken translational invariance in the SCTDHA formalism as presented in \cite{Nieuwkerk2018b}. Combined with the sector coupling, we find that inhomogeneities in the density can have weak but nontrivial effects on the amplitude of Josephson oscillations. This means that the trapping potential is likely to have an effect on the dynamics probed in experiment. In a forthcoming paper, we will present a study of the projected Hamiltonian (\ref{eq:micr_1d_ham}) in a microscopic analysis that takes a quadratic longitudinal potential into account. It would be interesting to combine such a microscopic approach with low-energy effective field theory calculations in the presence of such a quadratic trapping potential. However, the calculations using the box potential presented here may gain additional relevance when more experiments using a box potential, such as Refs. \cite{Rauer2018Box,Tajik2019}, are performed.


\section*{Acknowledgements}

We are grateful to J\"{o}rg Schmiedmayer, Thomas Schweigler and Marine Pigneur for
stimulating discussions and to the Erwin Schr\"odinger
International Institute for Mathematics and Physics for hospitality
and support during the programme on \emph{Quantum Paths}. This work
was supported by the EPSRC under grant EP/S020527/1 and YDvN is 
supported by the Merton College Buckee Scholarship and the VSB, Muller
and Prins Bernhard Foundations.

\appendix

\section{Tensors occurring in \texorpdfstring{$H_{\mathrm{SCH}}(t)$}{HSCH}} 
\label{sec:tensors_occurring_in_HSCH}

The tensors occurring in $H_{\mathrm{SCH}}(t)$ as written in Eq. (\ref{eq:generia_quadratic_ham}) are given by
\begin{align}
    A &= \begin{pmatrix}
        \ddots&&\reflectbox{$\ddots$}&\vline& \ddots & &\reflectbox{$\ddots$}\\
        &v q \, \delta_{q,k} + 2 \Delta^{(1)}_{q,k}(t) \,u^{(a)}_{a,q\vphantom{k}}(0) u^{(a)}_{a,k}(0) &  &\vline&  & \Delta^{(2)}_{q,k}(t) \,u^{(a)^*}_{a,q\vphantom{k}}(0) w^{(s)}_{s,k}(0) &  \\
        \reflectbox{$\ddots$}&&\ddots &\vline& \reflectbox{$\ddots$} & & \ddots \\
        \hline
        \ddots &&\reflectbox{$\ddots$}&\vline&\ddots&  &\reflectbox{$\ddots$}\\
        & \Delta^{(2)}_{q,k}(t) \,u^{(a)}_{a,k}(0) w^{(s)^{*}}_{s,q\vphantom{k}}(0) &  &\vline&  & v q \, \delta_{q,k}& \\
        \reflectbox{$\ddots$}&&\ddots&\vline&\reflectbox{$\ddots$}&&\ddots
    \end{pmatrix},\notag\\
    B &= \begin{pmatrix}
        \ddots&&\reflectbox{$\ddots$}&\vline&\ddots& &\reflectbox{$\ddots$}\\
        &-\frac{v \pi}{L} \delta_{q,k}\delta_{q,0} + 2 \Delta^{(1)}_{q,k}(t) \,u^{(a)}_{a,q\vphantom{k}}(0) u^{(a)}_{a,k}(0) &  &\vline&  & \Delta^{(2)}_{q,k}(t) \,u^{(a)}_{a,q\vphantom{k}}(0) w^{(s)}_{s,k}(0) &\\
        \reflectbox{$\ddots$}&&\ddots &\vline& \reflectbox{$\ddots$} &  & \ddots \\
        \hline
        \ddots&&\reflectbox{$\ddots$} &\vline&\ddots&  & \reflectbox{$\ddots$} \\
        &\Delta^{(2)}_{q,k}(t) \,u^{(a)}_{a,k}(0) w^{(s)}_{s,q\vphantom{k}}(0) &  &\vline&  & -\frac{v \pi}{L} \delta_{q,k}\delta_{q,0}& \\
        \reflectbox{$\ddots$} &&\ddots&\vline& \reflectbox{$\ddots$} &&\ddots
    \end{pmatrix},\notag\\
    D &= \begin{pmatrix}
        \vdots \\
        \Gamma^{(1)}_{q}(t)u^{(a)}_{a,q}(0) \\
        \vdots \\
        \hline
        \vdots \\
        0 \\
        \vdots
    \end{pmatrix},\;\;\;\;
    E = \begin{pmatrix}
        \vdots \\
        0 \\
        \vdots \\
        \hline
        \vdots \\
        \Gamma^{(2)}_{q}(t)w^{(s)}_{s,q}(0) \\
        \vdots
    \end{pmatrix}.
\end{align}
The momentum indices $q,k$ run within the blocks demarcated by solid lines, whereas the sector indices $j=a,s$ change from one block to the other. The functions occurring above are defined via
\begin{align}
\Gamma^{(i)}_{q}(t) &= \int_{0}^{L} dx \, g^{(i)}(x,t) \cos \left( q x \right), \\
\Delta^{(i)}_{q,k}(t) &= \int_{0}^{L} dx \, h^{(i)}(x,t) \cos \left( q x \right) \cos \left( k x \right) = \frac{1}{2}\left( h^{(i)}_{q+k}(t) + h^{(i)}_{|q-k|}(t) \right) \\
h^{(i)}_{q}(t) &= \int_{0}^{L} dx \, h^{(i)}(x,t) \cos \left( q x \right).
\end{align}



\begin{thebibliography}{10}
\providecommand{\url}[1]{\texttt{#1}}
\providecommand{\urlprefix}{URL }
\expandafter\ifx\csname urlstyle\endcsname\relax
  \providecommand{\doi}[1]{doi:\discretionary{}{}{}#1}\else
  \providecommand{\doi}{doi:\discretionary{}{}{}\begingroup
  \urlstyle{rm}\Url}\fi
\providecommand{\eprint}[2][]{\url{#2}}

\bibitem{Rigol2007}
M.~Rigol, V.~Dunjko, V.~Yurovsky and M.~Olshanii,
\newblock \emph{Relaxation in a completely integrable many-body quantum system:
  An ab initio study of the dynamics of the highly excited states of 1d lattice
  hard-core bosons},
\newblock Phys. Rev. Lett. \textbf{98}, 050405 (2007),
\newblock \doi{10.1103/PhysRevLett.98.050405}.

\bibitem{CalabreseCardy2007}
P.~Calabrese and J.~Cardy,
\newblock \emph{Quantum quenches in extended systems},
\newblock Journal of Statistical Mechanics: Theory and Experiment
  \textbf{2007}(06), P06008 (2007),
\newblock \doi{10.1088/1742-5468/2007/06/P06008}.

\bibitem{Polkovnikov2011}
A.~Polkovnikov, K.~Sengupta, A.~Silva and M.~Vengalattore,
\newblock \emph{Colloquium: Nonequilibrium dynamics of closed interacting
  quantum systems},
\newblock Rev. Mod. Phys. \textbf{83}, 863 (2011),
\newblock \doi{10.1103/RevModPhys.83.863}.

\bibitem{EFreview}
F.~H.~L. Essler and M.~Fagotti,
\newblock \emph{Quench dynamics and relaxation in isolated integrable quantum
  spin chains},
\newblock Journal of Statistical Mechanics: Theory and Experiment
  \textbf{2016}(6), 064002 (2016),
\newblock \doi{10.1088/1742-5468/2016/06/064002}.

\bibitem{Vidmar2016}
L.~Vidmar and M.~Rigol,
\newblock \emph{Generalized gibbs ensemble in integrable lattice models},
\newblock Journal of Statistical Mechanics: Theory and Experiment
  \textbf{2016}(6), 064007 (2016),
\newblock \doi{10.1088/1742-5468/2016/06/064007}.

\bibitem{DAlessio2016}
L.~D'Alessio, Y.~Kafri, A.~Polkovnikov and M.~Rigol,
\newblock \emph{From quantum chaos and eigenstate thermalization to statistical
  mechanics and thermodynamics},
\newblock Advances in Physics \textbf{65}(3), 239 (2016),
\newblock \doi{10.1080/00018732.2016.1198134}.

\bibitem{Gogolin2016}
C.~Gogolin and J.~Eisert,
\newblock \emph{Equilibration, thermalisation, and the emergence of statistical
  mechanics in closed quantum systems},
\newblock Reports on Progress in Physics \textbf{79}(5), 056001 (2016),
\newblock \doi{10.1088/0034-4885/79/5/056001}.

\bibitem{CalabreseCardy2016}
P.~Calabrese and J.~Cardy,
\newblock \emph{Quantum quenches in $1+1$ dimensional conformal field
  theories},
\newblock Journal of Statistical Mechanics: Theory and Experiment
  \textbf{2016}(6), 064003 (2016),
\newblock \doi{10.1088/1742-5468/2016/06/064003}.

\bibitem{CalabreseRev2018}
P.~Calabrese,
\newblock \emph{Entanglement and thermodynamics in non-equilibrium isolated
  quantum systems},
\newblock Physica A: Statistical Mechanics and its Applications \textbf{504},
  31  (2018),
\newblock \doi{10.1016/j.physa.2017.10.011},
\newblock Lecture Notes of the 14th International Summer School on Fundamental
  Problems in Statistical Physics.

\bibitem{Ketterle2001}
A.~G\"orlitz, J.~M. Vogels, A.~E. Leanhardt, C.~Raman, T.~L. Gustavson, J.~R.
  Abo-Shaeer, A.~P. Chikkatur, S.~Gupta, S.~Inouye, T.~Rosenband and
  W.~Ketterle,
\newblock \emph{Realization of {B}ose{-E}instein {C}ondensates in lower
  dimensions},
\newblock Phys. Rev. Lett. \textbf{87}, 130402 (2001),
\newblock \doi{10.1103/PhysRevLett.87.130402}.

\bibitem{Greiner2001}
M.~Greiner, I.~Bloch, O.~Mandel, T.~H{\"a}nsch and T.~Esslinger,
\newblock \emph{{B}ose{-E}instein condensates in {1D}- and {2D} optical
  lattices},
\newblock Applied Physics B \textbf{73}(8), 769 (2001),
\newblock \doi{10.1007/s003400100744}.

\bibitem{Kinoshita2004}
T.~Kinoshita, T.~Wenger and D.~S. Weiss,
\newblock \emph{Observation of a one-dimensional {T}onks-{G}irardeau gas},
\newblock Science \textbf{305}(5687), 1125 (2004),
\newblock \doi{10.1126/science.1100700}.

\bibitem{Schumm2005}
T.~Schumm, S.~Hofferberth, L.~M. Andersson, S.~Wildermuth, S.~Groth,
  I.~Bar-Joseph, J.~Schmiedmayer and P.~Kruger,
\newblock \emph{Matter-wave interferometry in a double well on an atom chip},
\newblock Nat Phys \textbf{1}(1), 57 (2005),
\newblock \doi{10.1038/nphys125}.

\bibitem{Hofferberth2007}
S.~{Hofferberth}, I.~{Lesanovsky}, B.~{Fischer}, T.~{Schumm} and
  J.~{Schmiedmayer},
\newblock \emph{{Non-equilibrium coherence dynamics in one-dimensional Bose
  gases}},
\newblock Nature \textbf{449}, 324 (2007),
\newblock \doi{10.1038/nature06149}.

\bibitem{Trotzky2012}
S.~Trotzky, Y.-A. Chen, A.~Flesch, I.~P. McCulloch, U.~Schollw{\"o}ck,
  J.~Eisert and I.~Bloch,
\newblock \emph{Probing the relaxation towards equilibrium in an isolated
  strongly correlated one-dimensional bose gas},
\newblock Nature Physics \textbf{8}, 325 EP  (2012),
\newblock \doi{10.1038/nphys2232}.

\bibitem{Cheneau2012}
M.~Cheneau, P.~Barmettler, D.~Poletti, M.~Endres, P.~Schau{\ss}, T.~Fukuhara,
  C.~Gross, I.~Bloch, C.~Kollath and S.~Kuhr,
\newblock \emph{Light-cone-like spreading of correlations in a quantum
  many-body system},
\newblock Nature \textbf{481}, 484 EP  (2012),
\newblock \doi{10.1038/nature10748}.

\bibitem{Gring2012}
M.~Gring, M.~Kuhnert, T.~Langen, T.~Kitagawa, B.~Rauer, M.~Schreitl, I.~Mazets,
  D.~A. Smith, E.~Demler and J.~Schmiedmayer,
\newblock \emph{Relaxation and prethermalization in an isolated quantum
  system},
\newblock Science \textbf{337} (6100), 1318 (2012),
\newblock \doi{10.1126/science.1224953}.

\bibitem{Langen2013}
T.~Langen, R.~Geiger, M.~Kuhnert, B.~Rauer and J.~Schmiedmayer,
\newblock \emph{Local emergence of thermal correlations in an isolated quantum
  many-body system},
\newblock Nat Phys \textbf{9}(10), 640 (2013),
\newblock \doi{10.1038/nphys2739},
\newblock Letter.

\bibitem{Langen2015}
T.~Langen, S.~Erne, R.~Geiger, B.~Rauer, T.~Schweigler, M.~Kuhnert,
  W.~Rohringer, I.~E. Mazets, T.~Gasenzer and J.~Schmiedmayer,
\newblock \emph{Experimental observation of a generalized gibbs ensemble},
\newblock Science \textbf{348}(6231), 207 (2015),
\newblock \doi{10.1126/science.1257026}.

\bibitem{Kaufman2016}
A.~M. Kaufman, M.~E. Tai, A.~Lukin, M.~Rispoli, R.~Schittko, P.~M. Preiss and
  M.~Greiner,
\newblock \emph{Quantum thermalization through entanglement in an isolated
  many-body system},
\newblock Science \textbf{353}(6301), 794 (2016),
\newblock \doi{10.1126/science.aaf6725}.

\bibitem{Kuhnert2013}
M.~Kuhnert, R.~Geiger, T.~Langen, M.~Gring, B.~Rauer, T.~Kitagawa, E.~Demler,
  D.~Adu~Smith and J.~Schmiedmayer,
\newblock \emph{Multimode dynamics and emergence of a characteristic length
  scale in a one-dimensional quantum system},
\newblock Phys. Rev. Lett. \textbf{110}, 090405 (2013),
\newblock \doi{10.1103/PhysRevLett.110.090405}.

\bibitem{Smith2013}
D.~A. Smith, M.~Gring, T.~Langen, M.~Kuhnert, B.~Rauer, R.~Geiger, T.~Kitagawa,
  I.~Mazets, E.~Demler and J.~Schmiedmayer,
\newblock \emph{Prethermalization revealed by the relaxation dynamics of full
  distribution functions},
\newblock New Journal of Physics \textbf{15}(7), 075011 (2013).

\bibitem{Schweigler2017}
T.~Schweigler, V.~Kasper, S.~Erne, I.~Mazets, B.~Rauer, F.~Cataldini,
  T.~Langen, T.~Gasenzer, J.~Berges and J.~Schmiedmayer,
\newblock \emph{Experimental characterization of a quantum many-body system via
  higher-order correlations},
\newblock Nature \textbf{545}(7654), 323 (2017),
\newblock \doi{10.1038/nature22310}.

\bibitem{Pigneur2017}
M.~Pigneur, T.~Berrada, M.~Bonneau, T.~Schumm, E.~Demler and J.~Schmiedmayer,
\newblock \emph{Relaxation to a phase-locked equilibrium state in a
  one-dimensional bosonic josephson junction},
\newblock Phys. Rev. Lett. \textbf{120}, 173601 (2018),
\newblock \doi{10.1103/PhysRevLett.120.173601}.

\bibitem{Ketterle1997}
M.~R. Andrews, C.~G. Townsend, H.-J. Miesner, D.~S. Durfee, D.~M. Kurn and
  W.~Ketterle,
\newblock \emph{Observation of {I}nterference between {T}wo {B}ose
  {C}ondensates},
\newblock Science \textbf{275}(5300), 637 (1997),
\newblock \doi{10.1126/science.275.5300.637}.

\bibitem{Nieuwkerk2018}
Y.~D. van Nieuwkerk, J.~Schmiedmayer and F.~H.~L. Essler,
\newblock \emph{{Projective phase measurements in one-dimensional Bose gases}},
\newblock SciPost Phys. \textbf{5}, 46 (2018),
\newblock \doi{10.21468/SciPostPhys.5.5.046}.

\bibitem{cd-07}
R.~W. Cherng and E.~Demler,
\newblock \emph{Quantum noise analysis of spin systems realized with cold
  atoms},
\newblock New Journal of Physics \textbf{9}(1), 7 (2007).

\bibitem{Imambekov2007}
A.~{Imambekov}, V.~{Gritsev} and E.~{Demler},
\newblock \emph{{Fundamental noise in matter interferometers}}  (2007),
\newblock \eprint{https://arxiv.org/abs/cond-mat/0703766}.

\bibitem{lp-08}
A.~Lamacraft and P.~Fendley,
\newblock \emph{Order parameter statistics in the critical quantum ising
  chain},
\newblock Phys. Rev. Lett. \textbf{100}, 165706 (2008),
\newblock \doi{10.1103/PhysRevLett.100.165706}.

\bibitem{ia-13}
D.~A. Ivanov and A.~G. Abanov,
\newblock \emph{Characterizing correlations with full counting statistics:
  Classical ising and quantum $xy$ spin chains},
\newblock Phys. Rev. E \textbf{87}, 022114 (2013),
\newblock \doi{10.1103/PhysRevE.87.022114}.

\bibitem{sk-13}
Y.~Shi and I.~Klich,
\newblock \emph{Full counting statistics and the edgeworth series for matrix
  product states},
\newblock Journal of Statistical Mechanics: Theory and Experiment
  \textbf{2013}(05), P05001 (2013).

\bibitem{e-13}
V.~Eisler,
\newblock \emph{Universality in the full counting statistics of trapped
  fermions},
\newblock Phys. Rev. Lett. \textbf{111}, 080402 (2013),
\newblock \doi{10.1103/PhysRevLett.111.080402}.

\bibitem{k-14}
I.~Klich,
\newblock \emph{A note on the full counting statistics of paired fermions},
\newblock Journal of Statistical Mechanics: Theory and Experiment
  \textbf{2014}(11), P11006 (2014).

\bibitem{sp-17}
J.-M. St\'ephan and F.~Pollmann,
\newblock \emph{Full counting statistics in the haldane-shastry chain},
\newblock Phys. Rev. B \textbf{95}, 035119 (2017),
\newblock \doi{10.1103/PhysRevB.95.035119}.

\bibitem{CoEG17}
M.~Collura, F.~H.~L. Essler and S.~Groha,
\newblock \emph{Full counting statistics in the spin-1/2 heisenberg xxz chain},
\newblock Journal of Physics A: Mathematical and Theoretical \textbf{50}(41),
  414002 (2017).

\bibitem{nr-17}
K.~Najafi and M.~A. Rajabpour,
\newblock \emph{Full counting statistics of the subsystem energy for free
  fermions and quantum spin chains},
\newblock Phys. Rev. B \textbf{96}, 235109 (2017),
\newblock \doi{10.1103/PhysRevB.96.235109}.

\bibitem{hb-17}
S.~Humeniuk and H.~P. B\"uchler,
\newblock \emph{Full counting statistics for interacting fermions with
  determinantal quantum monte carlo simulations},
\newblock Phys. Rev. Lett. \textbf{119}, 236401 (2017),
\newblock \doi{10.1103/PhysRevLett.119.236401}.

\bibitem{lddz-15}
I.~Lovas, B.~D\'ora, E.~Demler and G.~Zar\'and,
\newblock \emph{Full counting statistics of time-of-flight images},
\newblock Phys. Rev. A \textbf{95}, 053621 (2017),
\newblock \doi{10.1103/PhysRevA.95.053621}.

\bibitem{bpc-18}
A.~Bastianello, L.~Piroli and P.~Calabrese,
\newblock \emph{Exact local correlations and full counting statistics for
  arbitrary states of the one-dimensional interacting bose gas},
\newblock Phys. Rev. Lett. \textbf{120}, 190601 (2018),
\newblock \doi{10.1103/PhysRevLett.120.190601}.

\bibitem{Groha18}
S.~Groha, F.~H.~L. Essler and P.~Calabrese,
\newblock \emph{Full counting statistics in the transverse field ising chain},
\newblock SciPost Phys. \textbf{4}, 43 (2018),
\newblock \doi{10.21468/SciPostPhys.4.6.043}.

\bibitem{Collura20}
M.~Collura and F.~H.~L. Essler,
\newblock \emph{How order melts after quantum quenches},
\newblock Phys. Rev. B \textbf{101}, 041110(R) (2020),
\newblock \doi{10.1103/PhysRevB.101.041110}.


\bibitem{Haldane1981}
F.~D.~M. Haldane,
\newblock \emph{Effective harmonic-fluid approach to low-energy properties of
  one-dimensional quantum fluids},
\newblock Phys. Rev. Lett. \textbf{47}, 1840 (1981),
\newblock \doi{10.1103/PhysRevLett.47.1840}.

\bibitem{Bistritzer2007}
R.~Bistritzer and E.~Altman,
\newblock \emph{Intrinsic dephasing in one-dimensional ultracold atom
  interferometers},
\newblock Proceedings of the National Academy of Sciences \textbf{104}(24),
  9955 (2007),
\newblock \doi{10.1073/pnas.0608910104}.

\bibitem{Kitagawa2010}
T.~Kitagawa, S.~Pielawa, A.~Imambekov, J.~Schmiedmayer, V.~Gritsev and
  E.~Demler,
\newblock \emph{Ramsey interference in one-dimensional systems: The full
  distribution function of fringe contrast as a probe of many-body dynamics},
\newblock Phys. Rev. Lett. \textbf{104}, 255302 (2010),
\newblock \doi{10.1103/PhysRevLett.104.255302}.

\bibitem{Kitagawa2011}
T.~Kitagawa, A.~Imambekov, J.~Schmiedmayer and E.~Demler,
\newblock \emph{The dynamics and prethermalization of one-dimensional quantum
  systems probed through the full distributions of quantum noise},
\newblock New Journal of Physics \textbf{13}(7), 073018 (2011),
\newblock \doi{10.1088/1367-2630/13/7/073018}.

\bibitem{Albiez2005}
M.~Albiez, R.~Gati, J.~F\"olling, S.~Hunsmann, M.~Cristiani and M.~K.
  Oberthaler,
\newblock \emph{Direct observation of tunneling and nonlinear self-trapping in
  a single bosonic josephson junction},
\newblock Phys. Rev. Lett. \textbf{95}, 010402 (2005),
\newblock \doi{10.1103/PhysRevLett.95.010402}.

\bibitem{Gati2006}
R.~Gati, M.~Albiez, J.~F{\"o}lling, B.~Hemmerling and M.~Oberthaler,
\newblock \emph{Realization of a single josephson junction for bose--einstein
  condensates},
\newblock Applied Physics B \textbf{82}(2), 207 (2006),
\newblock \doi{10.1007/s00340-005-2059-z}.

\bibitem{Levy2007}
S.~Levy, E.~Lahoud, I.~Shomroni and J.~Steinhauer,
\newblock \emph{The a.c. and d.c. josephson effects in a bose-einstein
  condensate},
\newblock Nature \textbf{449}, 579 EP  (2007),
\newblock \doi{10.1038/nature06186}.

\bibitem{Gritsev2007}
V.~Gritsev, A.~Polkovnikov and E.~Demler,
\newblock \emph{Linear response theory for a pair of coupled one-dimensional
  condensates of interacting atoms},
\newblock Phys. Rev. B \textbf{75}, 174511 (2007),
\newblock \doi{10.1103/PhysRevB.75.174511}.

\bibitem{EsslerKonik2005}
F.~H. Essler and R.~M. Konik,
\newblock \emph{Applications of Massive Integrable Quantum Field Theories to
  Problems in Condensed Matter Physics}, pp. 684--830,
\newblock World Scientific,
\newblock \doi{10.1142/9789812775344_0020} (2005).

\bibitem{Iucci2010}
A.~Iucci and M.~A. Cazalilla,
\newblock \emph{Quantum quench dynamics of the sine-gordon model in some
  solvable limits},
\newblock New Journal of Physics \textbf{12}(5), 055019 (2010).

\bibitem{Foini2015}
L.~Foini and T.~Giamarchi,
\newblock \emph{Nonequilibrium dynamics of coupled luttinger liquids},
\newblock Phys. Rev. A \textbf{91}, 023627 (2015),
\newblock \doi{10.1103/PhysRevA.91.023627}.

\bibitem{Foini2017}
L.~Foini and T.~Giamarchi,
\newblock \emph{Relaxation dynamics of two coherently coupled one-dimensional
  bosonic gases},
\newblock The European Physical Journal Special Topics \textbf{226}(12), 2763
  (2017),
\newblock \doi{10.1140/epjst/e2016-60383-x}.

\bibitem{Bertini2014}
B.~Bertini, D.~Schuricht and F.~H.~L. Essler,
\newblock \emph{Quantum quench in the sine-gordon model},
\newblock Journal of Statistical Mechanics: Theory and Experiment
  \textbf{2014}(10), P10035 (2014).

\bibitem{Schuricht2017}
A.~C. Cubero and D.~Schuricht,
\newblock \emph{Quantum quench in the attractive regime of the sine-gordon
  model},
\newblock Journal of Statistical Mechanics: Theory and Experiment
  \textbf{2017}(10), 103106 (2017).

\bibitem{Horvath2017}
D.~Horváth and G.~Takács,
\newblock \emph{Overlaps after quantum quenches in the sine-gordon model},
\newblock Physics Letters B \textbf{771}, 539  (2017),
\newblock \doi{https://doi.org/10.1016/j.physletb.2017.05.087}.

\bibitem{Horvath2018a}
D.~X. Horv{\'a}th, M.~Kormos and G.~Tak{\'a}cs,
\newblock \emph{Overlap singularity and time evolution in integrable quantum
  field theory},
\newblock Journal of High Energy Physics \textbf{2018}(8), 170 (2018),
\newblock \doi{10.1007/JHEP08(2018)170}.

\bibitem{Kormos2016}
M.~Kormos and G.~Zar\'and,
\newblock \emph{Quantum quenches in the sine-gordon model: A semiclassical
  approach},
\newblock Phys. Rev. E \textbf{93}, 062101 (2016),
\newblock \doi{10.1103/PhysRevE.93.062101}.

\bibitem{Pascu2017}
C.~Moca, M.~Kormos and G.~Zar\'and,
\newblock \emph{Hybrid semiclassical theory of quantum quenches in
  one-dimensional systems},
\newblock Phys. Rev. Lett. \textbf{119}, 100603 (2017),
\newblock \doi{10.1103/PhysRevLett.119.100603}.

\bibitem{James2018}
A.~J.~A. James, R.~M. Konik, P.~Lecheminant, N.~J. Robinson and A.~M. Tsvelik,
\newblock \emph{Non-perturbative methodologies for low-dimensional
  strongly-correlated systems: From non-abelian bosonization to truncated
  spectrum methods},
\newblock Reports on Progress in Physics \textbf{81}(4), 046002 (2018),
\newblock \doi{10.1088/1361-6633/aa91ea}.

\bibitem{Kukuljan2018}
I.~Kukuljan, S.~Sotiriadis and G.~Takacs,
\newblock \emph{Correlation functions of the quantum sine-gordon model in and
  out of equilibrium},
\newblock Phys. Rev. Lett. \textbf{121}, 110402 (2018),
\newblock \doi{10.1103/PhysRevLett.121.110402}.

\bibitem{Sotiriadis2018}
I.~Kukuljan, S.~Sotiriadis and G.~Takacs,
\newblock \emph{Correlation functions of the quantum sine-gordon model in and
  out of equilibrium},
\newblock Phys. Rev. Lett. \textbf{121}, 110402 (2018),
\newblock \doi{10.1103/PhysRevLett.121.110402}.

\bibitem{Pigneur2019thesis}
M.~Pigneur,
\newblock \emph{Relaxation to a phase-locked equilibrium state in a
  one-dimensional bosonic Josephson junction},
\newblock Ph.D. thesis (2019).

\bibitem{schweigler2019correlations}
T.~Schweigler,
\newblock \emph{Correlations and dynamics of tunnel-coupled one-dimensional
  Bose gases},
\newblock Ph.D. thesis (2019), \eprint{https://arxiv.org/abs/1908.00422}.

\bibitem{Pigneur2018}
M.~Pigneur and J.~Schmiedmayer,
\newblock \emph{Analytical pendulum model for a bosonic josephson junction},
\newblock Phys. Rev. A \textbf{98}, 063632 (2018),
\newblock \doi{10.1103/PhysRevA.98.063632}.

\bibitem{Nieuwkerk2018b}
Y.~D. van Nieuwkerk and F.~H.~L. Essler,
\newblock \emph{Self-consistent time-dependent harmonic approximation for the
  sine-gordon model out of equilibrium},
\newblock Journal of Statistical Mechanics: Theory and Experiment
  \textbf{2019}(8), 084012 (2019),
\newblock \doi{10.1088/1742-5468/ab3579}.

\bibitem{Horvath2018}
D.~X. Horv\'ath, I.~Lovas, M.~Kormos, G.~Tak\'acs and G.~Zar\'and,
\newblock \emph{Nonequilibrium time evolution and rephasing in the quantum
  sine-gordon model},
\newblock Phys. Rev. A \textbf{100}, 013613 (2019),
\newblock \doi{10.1103/PhysRevA.100.013613}.

\bibitem{Rauer2018Box}
B.~Rauer, S.~Erne, T.~Schweigler, F.~Cataldini, M.~Tajik and J.~Schmiedmayer,
\newblock \emph{Recurrences in an isolated quantum many-body system},
\newblock Science  (2018),
\newblock \doi{10.1126/science.aan7938}.

\bibitem{Tajik2019}
M.~Tajik, B.~Rauer, T.~Schweigler, F.~Cataldini, J.~{a}o Sabino, F.~S.
  M{\o}ller, S.-C. Ji, I.~E. Mazets and J.~Schmiedmayer,
\newblock \emph{Designing arbitrary one-dimensional potentials on an atom
  chip},
\newblock Opt. Express \textbf{27}(23), 33474 (2019),
\newblock \doi{10.1364/OE.27.033474}.

\bibitem{Olshanii1998}
M.~Olshanii,
\newblock \emph{Atomic scattering in the presence of an external confinement
  and a gas of impenetrable bosons},
\newblock Phys. Rev. Lett. \textbf{81}, 938 (1998),
\newblock \doi{10.1103/PhysRevLett.81.938}.

\bibitem{TBP}
Y.~D. van Nieuwkerk and F.~H.~L. Essler, in preparation.

\bibitem{Cazalilla2004}
M.~A. Cazalilla,
\newblock \emph{Bosonizing one-dimensional cold atomic gases},
\newblock Journal of Physics B: Atomic, Molecular and Optical Physics
  \textbf{37}(7), S1 (2004),
\newblock \doi{10.1088/0953-4075/37/7/051}.

\bibitem{Hofferberth2008}
S.~{Hofferberth}, I.~{Lesanovsky}, T.~{Schumm}, A.~{Imambekov}, V.~{Gritsev},
  E.~{Demler} and J.~{Schmiedmayer},
\newblock \emph{{Probing quantum and thermal noise in an interacting many-body
  system}},
\newblock Nature Physics \textbf{4}, 489 (2008),
\newblock \doi{10.1038/nphys941},
\newblock \eprint{0710.1575}.

\bibitem{Schaff2015}
J.-F. {Schaff}, T.~{Langen} and J.~{Schmiedmayer},
\newblock \emph{Interferometry with atoms},
\newblock La Rivista del Nuovo Cimento  (2014),
\newblock \doi{10.1393/ncr/i2014-10105-7}.

\bibitem{Boyanovsky1998}
D.~Boyanovsky, F.~Cooper, H.~J. de~Vega and P.~Sodano,
\newblock \emph{Evolution of inhomogeneous condensates: Self-consistent
  variational approach},
\newblock Phys. Rev. D \textbf{58}, 025007 (1998),
\newblock \doi{10.1103/PhysRevD.58.025007}.

\bibitem{Sotiriadis2010}
S.~{Sotiriadis} and J.~{Cardy},
\newblock \emph{{Quantum quench in interacting field theory: A self-consistent
  approximation}},
\newblock Phys. Rev. B \textbf{81}(13), 134305 (2010),
\newblock \doi{10.1103/PhysRevB.81.134305},
\newblock \eprint{1002.0167}.

\bibitem{Lerose19}
A.~ Lerose, B. Zunkovic, J. Marino, A. Gambassi and A. Silva,
\newblock\emph{Impact of non-equilibrium fluctuations on pre-thermal
  dynamical phase transitions in long-range interacting spin chains},
\newblock Phys. Rev. B{\bf 99}, 045128 (2019),
\newblock \doi{10.1103/PhysRevB.99.045128}.

\end{thebibliography}

\end{document}